%% file: paper.tex
\documentclass{egpubl}
\usepackage{egsgp2026}
 
\SpecialIssuePaper         % uncomment for final version of Computer Graphics Forum, special issue
\CGFccby
\usepackage[T1]{fontenc}
\usepackage{dfadobe}

\usepackage{cite}  % comment out for biblatex with backend=biber
\BibtexOrBiblatex
\electronicVersion
\PrintedOrElectronic
\ifpdf \usepackage[pdftex]{graphicx} \pdfcompresslevel=9
\else \usepackage[dvips]{graphicx} \fi

\usepackage{egweblnk} 
\input{packages_and_commands}
\title{Phong-Rodrigues Extrinsic Vector-Field Processing}
\author[FP2-1031]{FP2-1031}
\author[H. Liu, O. Stein, A. Vaxman, M. Ben-Chen, \& M. Kazhdan]
{\parbox{\textwidth}{\centering
Hongyi Liu$^1$\orcid{0000-0002-5908-4822}
\qquad
Oded Stein$^2$\orcid{0000-0001-9741-3175}
\qquad
Amir Vaxman$^3$\orcid{0000-0001-6998-6689}
\qquad
Mirela Ben-Chen$^2$\orcid{0000-0002-1732-2327}
\qquad
Misha Kazhdan$^1$\orcid{0000-0002-6904-2167}
}
\\
{\parbox{\textwidth}{\centering $^1$Johns Hopkins University, USA\quad$^2$Technion - Israel Institute of Technology, Israel\quad$^3$University of Edinburgh, UK}
}
}
\begin{document}

%uncomment for using teaser
\teaser{
    \centering
    \includegraphics[width=0.25\textwidth]{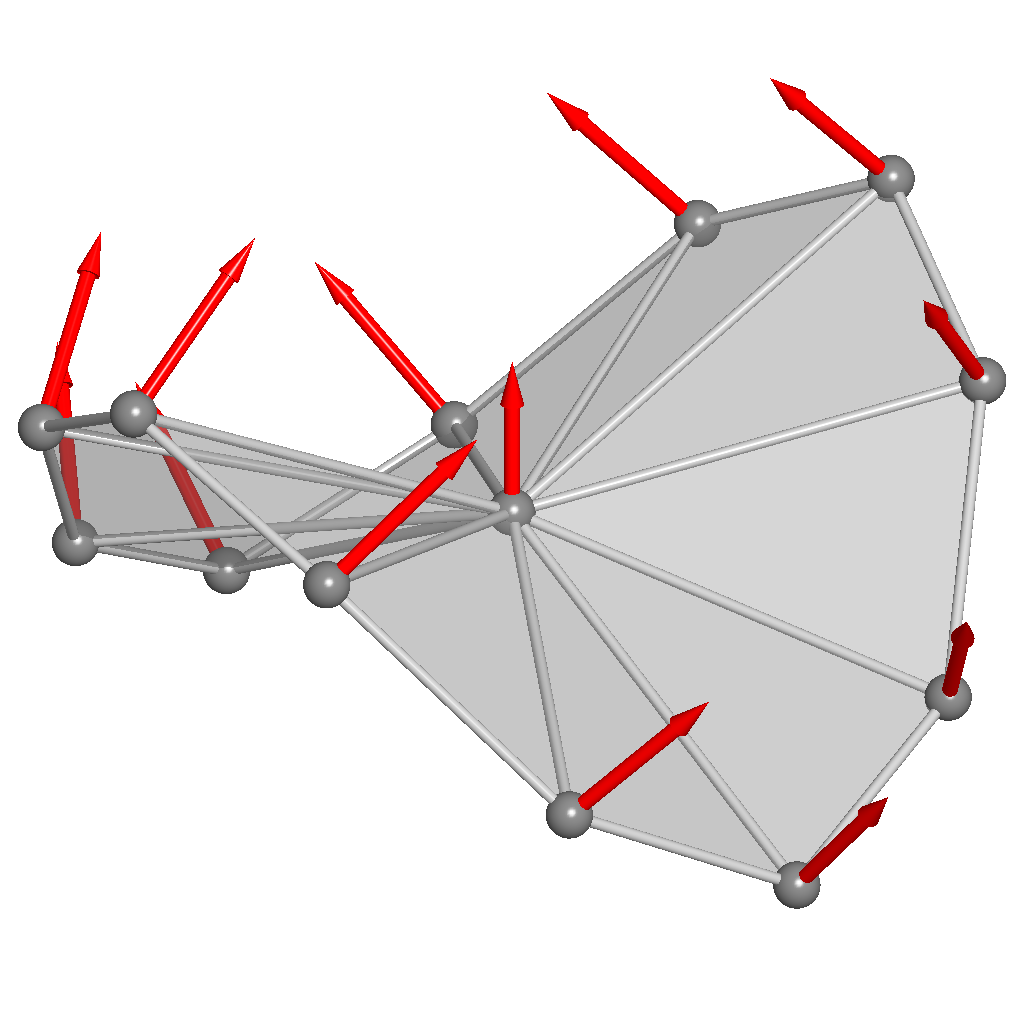}
    \includegraphics[width=0.25\textwidth]{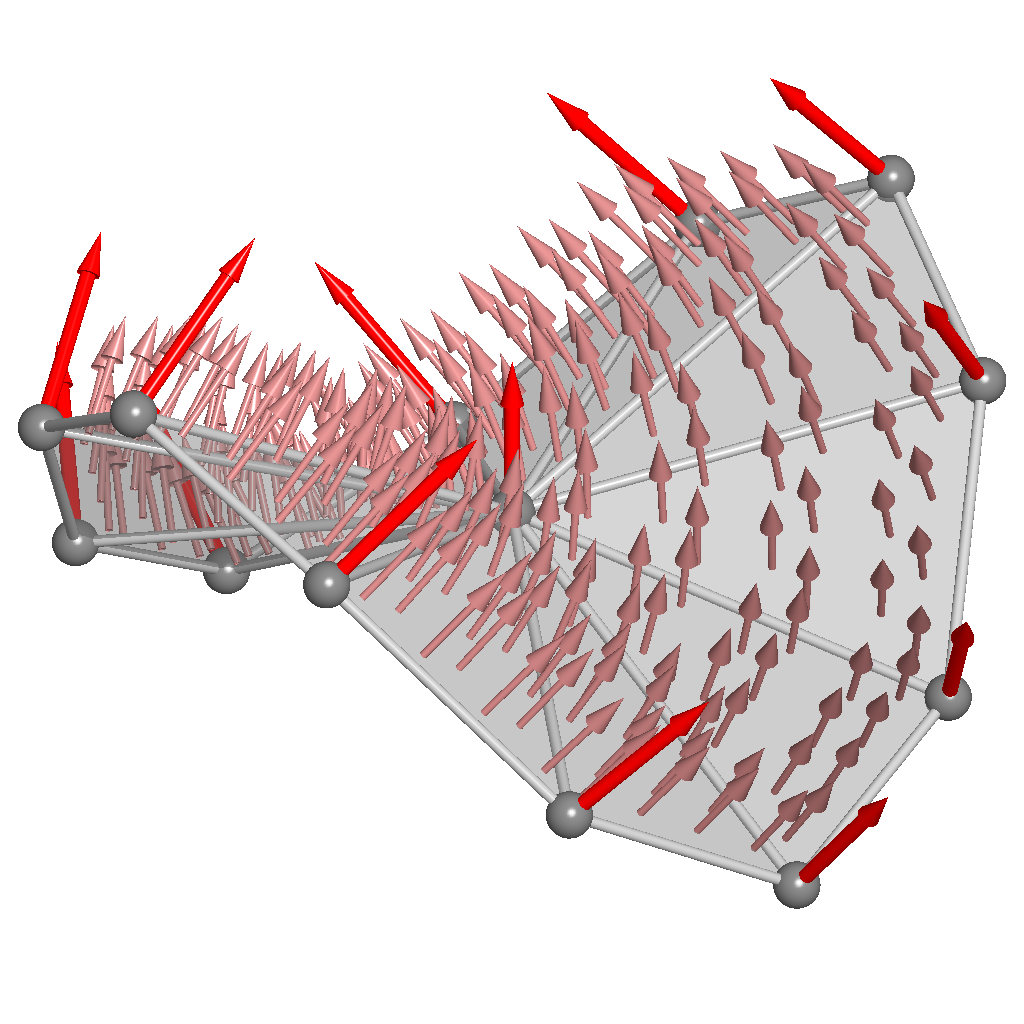}
    \includegraphics[width=0.25\textwidth]{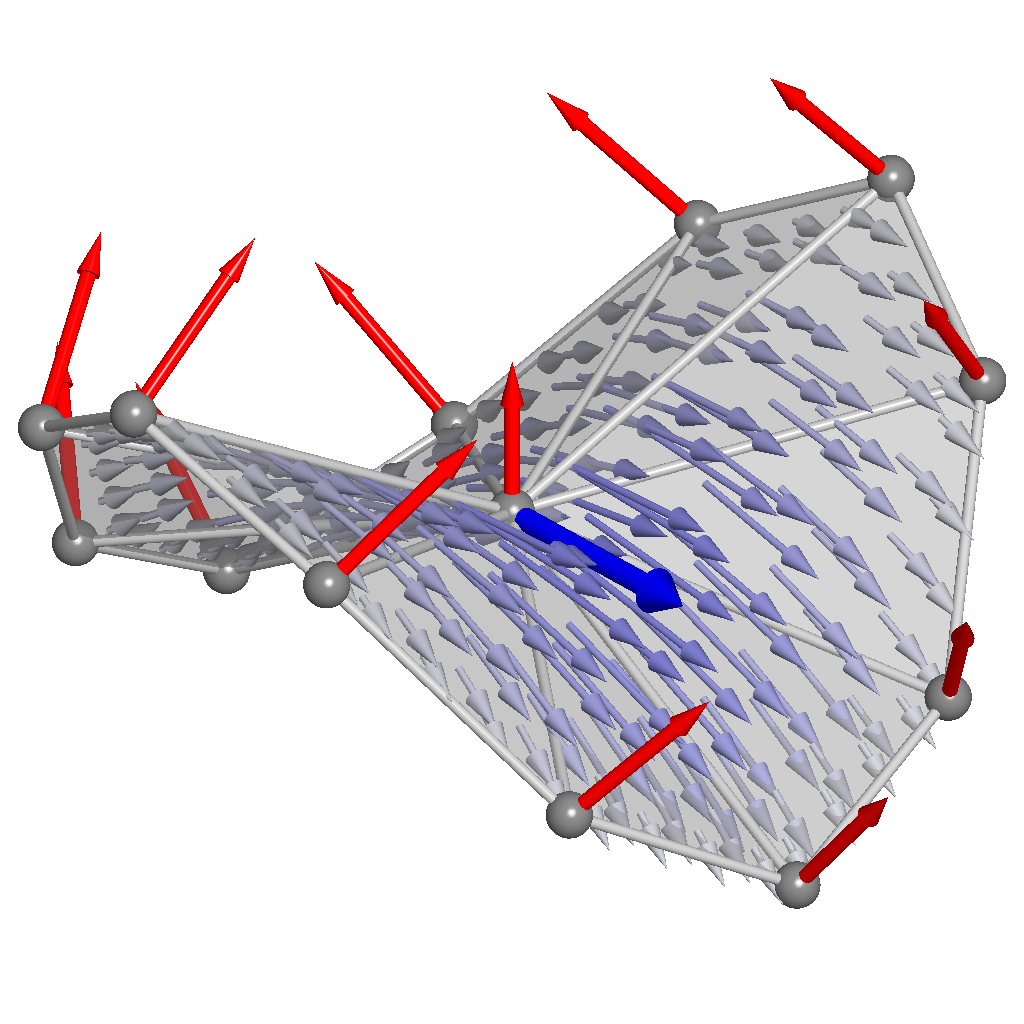}
    \caption{To define a continuous basis for tangent vector-fields over a triangle mesh, we start with a triangle mesh with per-vertex normals (left). Using Phong interpolation, we obtain a continuous normal field over the mesh (center). Then, given a tangent vector at a vertex (dark blue), we extend it into the interior of the incident triangles by (1)~transporting the vertex's tangent using the Rodrigues rotation taking the vertex's normal to the interior point's normal, and (2) scaling the transported tangent by the vertex's barycentric coordinate.}
    \label{fig:teaser}
}

\maketitle
%-------------------------------------------------------------------------
 \begin{abstract}
\input{Sections/00_abstract}


\end{abstract}  

%-------------------------------------------------------------------------
\section{Introduction}
\label{sec:introduction}
\input{Sections/01_introduction}

\section{Related Work}
\label{sec:related_work}
\input{Sections/02_related_work}
\section{Background}
\label{sec:background}
\input{Sections/03_approach}

\section{Discretization}
\label{sec:implementation}
\input{Sections/04_implementation}

\section{Energies and Operators}
\label{sec:energies_and_operators}
\input{Sections/05_energies}

\section{Topology and N-Fields}
\label{sec:topology}
\input{Sections/06_topology}

\section{Evaluation}
\label{sec:evaluation}
\input{Sections/07_evaluation}

\section{Conclusion and Future Directions}
\label{sec:discussion}

%\MB{this section reads weird - we need a proper finishing paragraph} \MK{Looking forward to seeing what Amir puts down.}

\input{Sections/08_discussion}

\begin{comment}
\section{Questions \MB{remove this}}
\begin{itemize}
\item Should we adjust mesh size when comparing to Stein~\etal, given that their approach has DoFs at edges rather than vertices.
\item For sparse vector-field interpolation, should we be comparing to something else?
\item For vector heat, should we be comparing to something else?
\item What other discretizations should we be considering?
\item Should we try to compute the Hodge and Killing energies, as well as the bracket for the Whitney basis using the point-wise evaluation of the Whitney basis functions? (This would give two definitions of a Hodge energy, one associated with the structure-preserving/DEC formulation, and one associated with the Galerkin interpretation. Not clear that these should be the same.) 
\end{itemize}
\end{comment}

\section*{Acknowledgements}
Mirela Ben-Chen acknowledges the support of the Israel Science Foundation (grant No. 1073/21).

\bibliographystyle{eg-alpha-doi} 
\bibliography{paper}      
%\printbibliography

\appendix
\begin{figure*}
    \centering
    \includegraphics[width=\linewidth]{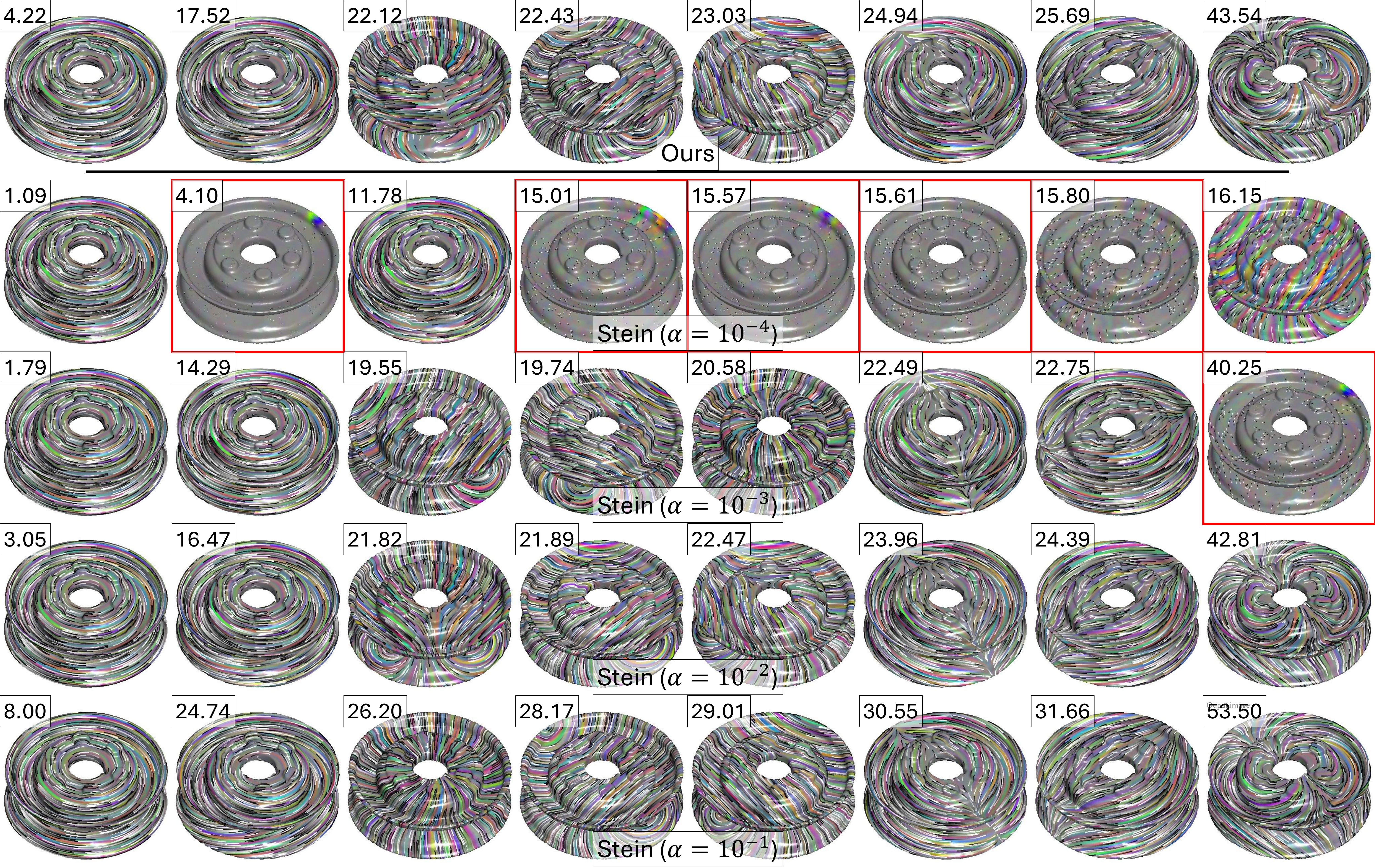}
    \caption{Spectra of the Killing energy: Comparing the smallest eigenvectors obtained using our approach (top row) with \revision{those} obtained Stein~\etal's method (second through fifth rows), with different regularization weights $\alpha$. Spurious minima are highlighted in red.}
    \label{fig:stein}
\end{figure*}

\revision
{
\section{Coordinate-Free Decomposition of Endomorphisms}
\label{a:endomorphism_decomposition}
\input{Sections/A3_appendix}
}

\section{Comparing Sparse Interpolation}
\label{a:sparse_interpolation}
\input{Sections/A2_appendix}

\section{Bracket Comparison}
\label{a:bracket}
\input{Sections/A1_appendix}

%-------------------------------------------------------------------------

\end{document}

%% file: packages_and_commands.tex
\usepackage{tikz}
\usetikzlibrary{chains}

\usepackage{amsmath}
\usepackage{amsfonts}
\usepackage{booktabs} % For formal tables
\usepackage{wrapfig}
\usepackage{subfigure}
\usepackage{textgreek}
\usepackage{pgfplots}
\usepackage{svg}
\usepackage{siunitx}
\usepackage{soul}
\usepackage{cleveref}
\usepackage{listings}
\usepackage{verbatim}
\usepackage{bm}
\usepackage{multirow}
\usepackage{enumitem}
\usepackage{adjustbox}
\usepackage{tikz}
\usepackage{nccmath}
\usetikzlibrary{matrix,positioning}
\usepackage{algpseudocode}
\usepackage{comment}
\usepackage{xspace}
\usepackage[scr=boondox,  % heavily sloped
            cal=esstix]   % slightly sloped
           {mathalpha}

\usepackage[ruled]{algorithm2e} % For algorithms
\crefname{algocf}{alg.}{algs.}
\Crefname{algocf}{Algorithm}{Algorithms}

\usepackage{bm}

\newtheorem{remark}{Remark}

\newcommand{\R}{{\mathbb R}}

\newcommand{\eqn}[1]{Eq.~\ref{#1}}
\newcommand{\fig}[1]{Fig.~\ref{#1}}
\newcommand{\tab}[1]{Tab.~\ref{#1}}
\renewcommand{\sec}[1]{Sec.~\ref{#1}}
\newcommand{\alg}[1]{Alg.~\ref{#1}}
\newcommand{\app}[1]{App.~\ref{#1}}

\DeclareMathOperator*{\argmin}{arg\!\,min}

\newcommand{\oldstuff}[1]{}

\newcommand{\revision}[1]{#1}

\newcommand{\etal}{\textit{et~al.}\xspace}

\makeatletter
\DeclareFontFamily{OMX}{MnSymbolE}{}
\DeclareSymbolFont{MnLargeSymbols}{OMX}{MnSymbolE}{m}{n}
\SetSymbolFont{MnLargeSymbols}{bold}{OMX}{MnSymbolE}{b}{n}
\DeclareFontShape{OMX}{MnSymbolE}{m}{n}{
    <-6>  MnSymbolE5
   <6-7>  MnSymbolE6
   <7-8>  MnSymbolE7
   <8-9>  MnSymbolE8
   <9-10> MnSymbolE9
  <10-12> MnSymbolE10
  <12->   MnSymbolE12
}{}
\DeclareFontShape{OMX}{MnSymbolE}{b}{n}{
    <-6>  MnSymbolE-Bold5
   <6-7>  MnSymbolE-Bold6
   <7-8>  MnSymbolE-Bold7
   <8-9>  MnSymbolE-Bold8
   <9-10> MnSymbolE-Bold9
  <10-12> MnSymbolE-Bold10
  <12->   MnSymbolE-Bold12
}{}

\let\llangle\@undefined
\let\rrangle\@undefined
\DeclareMathDelimiter{\llangle}{\mathopen}%
                     {MnLargeSymbols}{'164}{MnLargeSymbols}{'164}
\DeclareMathDelimiter{\rrangle}{\mathclose}%
                     {MnLargeSymbols}{'171}{MnLargeSymbols}{'171}
\makeatother

\newcommand{\T}{{\mathbb{T}}}

\newcommand{\asym}[1]{\mathcal{Curl}(#1)}
\newcommand{\scalar}[1]{\mathcal{Div}(#1)}
\newcommand{\noscalar}[1]{\overline{\mathcal{Holom}}(#1)}

%% file: Sections/00_abstract.tex
We introduce a new extrinsic discretization of tangent vector fields on triangle meshes that is continuous, with bounded derivatives that are continuous almost everywhere, supporting pointwise evaluation and integration of differential operators. We achieve this by building a continuous normal field over the mesh via Phong interpolation and using minimal Rodrigues rotations to transport vertex-based tangent vectors into triangle interiors. Unlike most existing discretizations, which typically sacrifice either continuity or the ability to evaluate derivatives pointwise, our approach supports both. 
Because it is pointwise evaluatable, and using the fact that the covariant derivative can be decomposed into its symmetric, antisymmetric, and scalar components, our discretization supports the construction of standard vector-field processing operators including the connection and Hodge Laplacians, Killing energy, divergence, curl, and the Lie bracket. This framework provides a simple and practical finite-element formulation for vector-field processing on meshes, supporting both integration-based operators and pointwise queries. To our knowledge, ours is the first discretization that jointly enables extrinsic continuous vector fields, bounded derivatives, and pointwise evaluation of this collection of operators.

%% file: Sections/01_introduction.tex
Vector-field processing is an essential step in a broad class of geometry-processing applications, supporting the study of deformations of signals over a mesh and has been a corner-stone in a large body of work, including frame design, deformation, parallel transport, and optical flow. An essential first step is defining a discretization of vector-fields, enabling their representation, the study of their properties (e.g., smoothness), and the formulation of energies that have desired vector-fields as their minima.

A key challenge in discretizing vector-fields, and in particular in supporting the study of vector-field smoothness, is that the ``natural'' definition of vector-fields would represent them by values associated with faces, where a tangent space can be defined as the space perpendicular to the face normal. However, such representations are inherently discontinuous, due to the piecewise-constant triangle normals, resulting in unbounded derivatives across edges and making smoothness energies challenging to define.
%\MB{don't we claim at some point that for integration it's fine things are discontinuous across edges?} \MK{The integrand can be discontinuous, but if we need to differentiate it (e.g. to compute the covariant derivative) then we're in trouble.}

These challenges have typically been addressed in one of two ways: Discrete Exterior Calculus (DEC) \cite{Hirani:PhDThesis:2003} has been used to define discretizations of vector-fields and associated smoothness energies \textit{without} explicitly constructing a basis. It provides a way to optimize vector-fields but only \revision{supports} evaluation at specific locations and along specific directions. The second approach is to use a discontinuous vector-field basis (e.g. N\'{e}d\'{e}lec/Whitney, Raviart-Thomas, or Crouzeix-Raviart \cite{Hiptmair:AMS:1999,Gatica:SIMFEM:2014,Stein:TOG:2020}). These support point-wise evaluation and differentiation, but discontinuity along edges results in only some differential operators begin well-formulated in this representation.

An exception to this are the representations proposed by Kn\"{o}ppel~\etal \shortcite{Knoppel:2013:TOG} and Liu~\etal~\cite{Liu:TOG:2016}. There, the authors propose extending an intrinsically represented tangent vector defined at a vertex into the incident triangles by a combination of parallel transport and linear scaling.

In our work we follow these approaches, replacing the intrinsic representation of tangent vectors with an explicit one, visualized in \fig{fig:teaser}. To this end we assume we are given a triangle mesh with per-vertex normals as input (left), which we extend to a continuous normal field over the mesh (center). Then, given a tangent vector at a vertex, we extend that vector to a continuous vector-field over the mesh (right). Concretely, given a point in the interior of the incident triangle we define the tangent vector at that point by (1)~using the Rodrigues rotation (taking the vertex's normal to the point's normal) to transport the tangent from the vertex to the point, and (2)~scaling the transported tangent by the component of the point's barycentric coordinate associated with the vertex.

%In addition to providing an extrinsic realization of \shortcite{Knoppel:2013:TOG,Liu:TOG:2016} that is consistent with per-vertex normals, our 
\revision{Our} approach provides a simple expression for the vector-field basis that supports point-wise evaluation and differentiation. This, in turn, make it possible to define a  covariant derivative that is point-wise evaluatable, enabling the construction of mass, stiffness (Hodge and Connection), and Killing energies. It also supports the point-wise evaluation of the Lie bracket of two vector-fields.

We demonstrate the efficacy of our approach in defining Connection, Hodge, and Killing energies, as well as in computing singularities, and estimating the Lie bracket. While subsets of these applications are supported by existing discretizations, the continuity of our vector-field basis makes it the first to support all of them.

%\MB{We compare our approach to xxx and show that yyy. In addition we show that our method is applicable to zzz}

%% file: Sections/02_related_work.tex
This work defines a continuous and pointwise-evaluatable extrinsic vector-field, together with its covariant derivative, for triangle meshes embedded in $\R^3$. This requires a choice of vector-field representation and a notion of transport between adjacent tangent spaces. Together these determine the covariant derivative, from which all other operators are derived. We review prior methods, focusing on continuity and pointwise-evaluability, and refer the reader to \cite{Vaxman2016, Goes2016} for comprehensive reviews of vector-field representations and applications. A summary comparison is provided in \tab{tab:related_work}.

\begin{table}[]
\centering
\begin{tabular}{l|cccr@{ / }l}
%\toprule
& \multirow{2}{*}{Rep.} & $C^0$ a.e. and & $C^1$ a.e. and & \multicolumn{2}{c}{\multirow{2}{*}{DoFs}} \\
& & bounded & bounded & \multicolumn{2}{c}{}\\
% Method & Rep. & \begin{tabular}[c]{@{}c@{}}$C^0$ a.e. and \\ bounded? \end{tabular}\begin{tabular}[c]
\midrule
\cite{Knoppel:2013:TOG} & In. & \checkmark & ? & 2&vertex\\
\cite{Azencot:TOG:2015} & Ex. & \checkmark & - & 3&vertex\\
\cite{Knoppel:2015:SPS} & In. & - & - & 2&vertex\\
\cite{Sharp:2019:TOG} & In. & - & - & 2&vertex\\
\cite{Liu:TOG:2016} & In. & \checkmark & \checkmark & 2&vertex\\
\cite{Whitney:GIT:1957} & In. & \checkmark & - & 1&edge\hspace{0.6em} \\
\cite{Stein:2020:CGF} & In. & \checkmark & - & 2&edge\hspace{0.6em} \\
\cite{Boksebeld2022}-PL & In. & \checkmark & - & 6&face\hspace{0.8em}  \\
Ours & Ex. & \checkmark & \checkmark & 2&vertex \\ \bottomrule
\end{tabular}
\caption{Comparison of vector-field discretization methods by intrinsic/extrinsic representation, almost-everywhere continuity and boundedness of the vector field and its covariant derivative, and degrees of freedom. For \cite{Boksebeld2022} we describe the piecewise-linear (PL) version. (Though the discretization of Kn\"oppel~\etal \cite{Knoppel:2013:TOG} should have an evaluatable covariant derivative, the lack of closed-form expression makes this difficult in practice.)}
\label{tab:related_work}
\end{table}

The most related work is the intrinsic discretization of Kn\"{o}ppel~\etal~\cite{Knoppel:2013:TOG} which defines a connection Laplacian on vertex-based tangent frames. % Each vertex carries a unit basis vector with angles rescaled so that the angle sum at each vertex is 2$\pi$, and each edge is assigned a transport rotation by the difference in angles that the two endpoint basis vectors make with the shared edge. 
These per-vertex frames are then extended into incident triangles via parallel transport along radii, yielding a connection Laplacian assembled in integrated form. Later works~\cite{Knoppel:2015:SPS, Sharp:2019:TOG} build on this representation, assembling the connection Laplacian purely from edge transport rotations without interpolating frames into triangle interiors, and are not pointwise-evaluatable.

As Liu~\etal~\cite{Liu:TOG:2016} noted, the connection Laplacian of Kn\"{o}ppel~\etal~\cite{Knoppel:2013:TOG} does not provide closed-form expressions for covariant derivatives, and first-order operators such as divergence and curl cannot be evaluated in their framework. Liu~\etal~\cite{Liu:TOG:2016} address this by additionally defining vertex-to-edge and vertex-to-triangle transition rotations, enabling explicit evaluation of the covariant derivatives. This requires introducing ``impulse rotations'' at \revision{ chart crossings.}
%, making the interpolated vector-field discontinuous extrinsically. %\MB{what are the main downsides of Liu16 compared to us then? Is it just the extrinsic thing as mentioned in the next paragraph? Do they have a FEM basis they can integrate against?} \HY{It is FEM. I think the paper has its own notion of defining covariant derivative, which is continuous intrinsically, but not extrinsically.} %However, this requires deliberately spreading Gaussian curvature around vertices, causing the connection to deviate from the canonical Levi-Civita connection, and Stokes' theorem to fail to hold locally. \HY{Does our method preserves these?}

Degrees of freedom can also be placed on faces or edges.  %Earlier methods use DEC~\cite{Desbrun:2008:SigCourse} directly by placing tangent directions on faces and encoding rotation angle on dual edges, but define no covariant derivative or connection Laplacians beyond connection itself \cite{Crane:2010:TCD}. 
DEC provides a framework for discrete differential operators on triangle meshes, which can be realized in the finite-elements framework using the Whitney basis~\cite{Wang2006, Fisher2007, Desbrun:2006:SigCourse}.
\revision{Crouzeix-Raviart bases, with two degrees of freedom per edge encoding the components parallel and perpendicular to the edge, have also been proposed \cite{Djerbetian2016,Stein:2020:CGF}.}
%Stein~\etal~\cite{Stein:2020:CGF} and Djerbetian~\cite{Djerbetian2016} adopt a Crouzeix-Raviart-type formulation with two degrees of freedom per edge, leveraging the fact that adjacent triangles are intrinsically flat when unfolded across the shared edges.
Boksebeld and Vaxman~\cite{Boksebeld2022} use a primal-dual decomposition, generalizing the connection Laplacian to higher order polynomials\revision{, with six degrees of freedom per face in the piecewise linear case}. However, \revision{these three formulations all yield discontinuous vector fields.}%the vector fields defined by these methods are all discontinuous.

%Custers and Vaxman \cite{Custers2020} represent piecewise-constant face-based vector-fields via their projections onto the halfedges of each triangle using a subdivision scheme, bridging the mixed finite-element representation with DEC approach. Later work also uses a primal-dual construction and generalized the connection Laplacian to higher order piecewise-polynomials with discontinuous fields \cite{Boksebeld2022}.

%In contrast to these methods that define discontinuous vector-fields, our vector-fields are continuous everywhere. To do so, we leverage the Phong representation of Boubekeur~\etal~\cite{Boubekeur:2008:TOG}, first defining a continuous normal field over the surface, and then defining a space of tangent vector-fields that is everywhere orthogonal to the normals. We will show that this definition also admits a pointwise evaluation of the covariant derivative.

While most methods are intrinsic, an extrinsic representation is given by Azencot~\etal~\cite{Azencot:TOG:2015} via differentiation of the ambient coordinates and projection onto the tangent plane. %, with directional derivatives represented in a truncated spectral basis. 
This builds on the operator view of vector-fields~\cite{Azencot:CGF:2013} where the Lie bracket arises naturally from operator composition.

\revision{Our extrinsic approach follows Knöppel et al. \cite{Knoppel:2013:TOG} and Liu et al. \cite{Liu:TOG:2016} in placing degrees of freedom on vertices and extending vector fields into triangle interiors, but differs in two key ways. First, instead of using the intrinsic tangent frames derived from the triangle mesh, we define per-point tangent planes extrinsically via Phong interpolation, decoupling the tangent bundle from the mesh geometry. This choice makes ambient-space continuity possible, in contrast to prior methods that tend to focus on intrinsic continuity.}
%Our approach is an \textit{extrinsic} analogue of the basis introduced by Kn\"{o}ppel~\etal~\cite{Knoppel:2013:TOG} and Liu~\etal~\cite{Liu:TOG:2016}. \revision{Rather than deriving tangent frames directly from the polyhedral geometry, we decouple the tangent bundle from the mesh by defining a per-point tangent plane via Phong interpolation.} 
%\revision{This extrinsic formulation also allows vector fields to be visualized directly as 3D vectors in the ambient space.}
%This is achieved by defining tangent spaces in terms of (continuous) Phong interpolation. 
\revision{Second, we define the covariant derivative by differentiating the extrinsic vector field in the ambient space and projecting onto the tangent plane, enabling the construction of a broad set of differential operators including the connection Laplacian, the Hodge Laplacian and the Lie bracket. This is similar to the approach of Azencot~\etal~\cite{Azencot:TOG:2015}.} %Azencot~\etal~\cite{Azencot:TOG:2015}, we define the covariant derivative by differentiating the extrinsic vector field and projecting onto the tangent plane. 
However, (1)~the vector-fields we use are perpendicular to the normals by construction, and (2)~our normals vary continuously over the surface. %\revision{Our discretization yields a continuous, pointwise-evaluatable extrinsic vector field with bounded derivatives, enabling the construction of a broad set of differential operators including connection Laplacian, Hodge Laplacian and the Lie bracket.} %\revision{The vector field defined by our method is continuous and pointwise-evaluatable everywhere.}

%% file: Sections/03_approach.tex
To define the covariant derivative and decompose it into its constituent parts, we leverage some basic linear algebra, which we briefly review here. We also recall the expression for the (minimal) Rodrigues rotation and introduce notation that we will use throughout the remainder of the discussion (\tab{tab:notation}).

%As our approach is based on defining a basis for function spaces on a mesh, composed of functions that are defined ``piecewise'' per triangle, our discussion will focus on a single embedded triangle. Standard finite element assembly is then used to combine the per-triangle quantities into global system matrices/vectors.

\subsection{Linear Algebra Review}
\label{ss:linear_algebra}
%\subsubsection*{Motivation}
We consider the covariant derivative of a vector-field $v$ -- a function assigning, to each point $p\in\mathcal{S}$ on the surface, an endomorphism on the tangent space at that point, $\nabla v|_p\in\hbox{End}(T_p\mathcal{S})$. Following earlier work, we compute an orthogonal decomposition of the covariant derivative into constituent components in order to extract the Killing, holomorphic, anti-holomorphic, etc. energies. %\revision{While this can be done by selecting an orthonormal basis and considering the skew-symmetric and symmetric components, we review how this is done in a coordinate-free manner.} %To this end, we require a notion of inner-product on the space of endomorphisms derived from the inner-product on the tangent space itself.

\revision{We review the linear algebra involved. Given a vector space $V$, we denote by $V^*$ its \textit{dual} -- the space of linear functions on $V$. Given vector spaces $V$ and $W$ and a homomorphism $L:V\rightarrow W$, we denote by $L^*:W^*\rightarrow V^*$ the dual homomorphism defined in terms of the pull-back/composition, with $L^*(\phi)\equiv\phi\circ L$ for all $\phi\in W^*$. Finally, we denote}
%We briefly review the linear algebra involved,  denoting
an inner-product space as $\{V,B:V\rightarrow V^*\}$, with $V$ the vector space, $V^*$ its dual, and $B$ the inner-product. Though the inner-product can be equivalently represented as a bilinear map, $B(v,w)\equiv [B(v)](w)$, its representation as a linear map (combined with the fact that inner-products are symmetric and positive-definite) makes the inverse $B^{-1}:V^*\rightarrow V$, well-defined.

\subsubsection*{Endomorphism Decomposition}
\revision{As in previous work \cite{deGoes:2014:CGF,Liu:TOG:2016}, we decompose the space of endomorphisms into three orthogonal subspaces and consider the projection of individual endomorphisms onto the subspaces. We review the decomposition using a coordinate-free formulation. The subspaces are:
\begin{itemize}
\item $\scalar{V}$ -- the space of scalar multiples: These describe infinitesimal isotropic scaling and, in the context of the covariant derivative, give the vector-field's divergence. (Represented with respect to any basis, these are scalar multiples of the identity.)
\item $\noscalar{V}$ -- the space of trace-free, self-adjoint operators: These describe infinitesimal, anisotropic, volume-preserving scaling and, in the context of the covariant derivative, give the vector-field's anti-holomorphic component. (Represented with respect to an orthonormal basis, these are symmetric matrices with vanishing trace.)
\item $\asym{v}$ -- the space of anti-self-adjoint operators: These describe infinitesimal rotations and, in the context of the covariant derivative, give the vector-field's curl. (Represented with respect to an orthonormal basis, these are skew-symmetric matrices.)
\end{itemize}
}
%We briefly summarize coordinate-free decomposition of the space of endomorphisms on an inner-product space $\{V,B:V\rightarrow V^*\}$. For a more detailed discussion of factorization of the space of 2-tensors, we refer the reader to the work of de~Goes~\etal \cite{deGoes:2014:CGF} and Liu~\etal~\cite{Liu:TOG:2016}.
\revision{Formally, an} endomorphism $L:V\rightarrow V$ is said to be \textit{self-adjoint} if it commutes with the inner-product:
$$B\circ L = L^* \circ B.$$
Analogously, an endomorphism $L:V\rightarrow V$ is \textit{anti-self-adjoint} if:
$$B\circ L = -L^* \circ B.$$

%the bilinear form $B_V\circ L:V\rightarrow V^*$ is symmetric:
%$$\left[(B_V\circ L)(v)\right](w)=\left[(B_V\circ L(w))\right](v),\quad\forall v,w\in V.$$
%Analogously, an endomorphism $L:V\rightarrow V$ is anti-self-adjoint if:
%$$\left[(B_V\circ L)(v)\right](w)=-\left[(B_V\circ L(w))\right](v),\quad\forall v,w\in V.$$

The subsets of self-adjoint and anti-self-adjoint endomorphisms form linear subspaces.
%, denoted $\sym{V}$ and $\asym{V}$ respectively.
Given an endomorphism $L:V\rightarrow V$, the orthogonal projections
%onto $\sym{V}$ and $\asym{V}$
are, respectively:
$$L \mapsto \left(\frac{L+B^{-1}\circ L^*\circ B}2\right)\quad\hbox{and}\quad
L \mapsto \left(\frac{L-B^{-1}\circ L^*\circ B}2\right).
$$
\revision{Represented with respect to an orthonormal base, this amounts to averaging the matrix with its transpose (resp. negative transpose), resulting in symmetric (resp. skew symmetric) matrices.}

The space
\revision{ of self-adjoint operators}
%$\sym{V}$
can be further decomposed into endomorphisms that are scalar multiples of the identity
%, $\scalar{V}$
and endomorphisms with vanishing trace.
%$\noscalar{V}$.
Given a self-adjoint endomorphism
\revision{$L\in\hbox{End}(V)$,}
%$L\in\sym{V}$,
the orthogonal projections onto the subspaces are, respectively:
$$
L\mapsto\frac{\hbox{tr}(L)}{\hbox{dim}(V)}\cdot\hbox{Id.}
\quad\hbox{and}\quad
L\mapsto L-\frac{\hbox{tr}(L)}{\hbox{dim}(V)}\cdot\hbox{Id.}
$$

\revision{In \app{a:endomorphism_decomposition} we show that these maps are orthogonal projections (with respect to the canonical inner-product on the space of endmorphisms $\hbox{End}(V)$ induced by the inner-product on $V$) providing an orthogonal decomposition the space of endomorphisms as:
$$\hbox{End}(V)=\scalar{V}\oplus\noscalar{V}\oplus\asym{V}.$$
}

%Thus, the partition of the space of endomorphisms:
%$$\hbox{End}(V)=\scalar{V}\oplus\noscalar{V}\oplus\asym{V}$$
%is an orthogonal decomposition of $\hbox{End}(V)$ with respect to the inner-product induced by $B$.

%We note that this factorization matches the one proposed by Liu and colleagues~\cite{Liu:TOG:2016}, \eqn~(4)

\subsection{Rodrigues Rotation}
Given (non-antipodal) unit vectors $v,w\in S^{d-1}$, the Rodrigues rotation formula gives the minimal-angle rotation taking $v$ to $w$:
$$R(v,w) \equiv \hbox{Id.}+(w\cdot v^\top - v\cdot w^\top) + \frac{(w\cdot v^\top - v\cdot w^\top)^2}{1+\langle v , w \rangle}\in\hbox{SO}(d).$$
This expression is singular only when $v$ and $w$ are antipodal, in which case the rotation by $180^\circ$ in \textit{any} plane containing $v$ will be minimal.) Otherwise, the coefficients of $R$ are smooth functions of $v$ and $w$, with derivatives that are readily computable.

\subsection{Notation}
\begin{table}
\small
\center{
  \begin{tabular}{l|l}
\toprule
  $R:S^2\times S^2\rightarrow \hbox{SO}(3)$ & Rodrigues rotation\\
  $v_i\in\R^3$ & Vertices of the triangle \\
  $n_i\in S^2$ & Normals of the triangle \\
  $n\in S^2$ & Normal defined by triangle $\{v_0,v_1,v_2\}$\\
  $\T\subset\R^2$ & Unit right triangle\\
  $S^2\subset\R^3$ & Unit sphere\\
  $T_p\T\simeq\R^2$ & Tangent space at $p\in\T$ \\
  $T_nS^2\subset\R^3$ & Tangent space at $n\in S^2$ \\
  $\hbox{End}(T_p\T)$ & The space of endomorphisms on $T_p\T$\\
  $\Phi:\T\rightarrow\R^3$ & Linear embedding of triangle $\{v_0,v_1,v_2\}$\\
  $d\Phi|_p:T_p\T\rightarrow T_nS^2$ & Differential of $\Phi$ at $p\in\T$\\
  $g|_p:T_p\T\rightarrow T_p^*\T$ & Metric on $T_p\T$ induced by $d\Phi|_p$\\
  $N:\T\rightarrow S^2$ & Gauss map defined by vertex normals\\
  $\widetilde{d\Phi}|_p:T_p\T\rightarrow T_{N(p)}S^2$ & Realization of $T_p\T$ in $T_{N(p)}S^2$ \\
  $\psi_i:\T\rightarrow\R$ & Hat basis\\
  $\mathbf{m}^\psi\in\R^{3\times3}$ & Element scalar mass matrix\\
  $\mathbf{s}^\psi\in\R^{3\times3}$ & Element scalar stiffness matrix\\
  $R_i:\T\rightarrow\hbox{SO}(3)$ & Transport from the $i$-th corner\\
  $\overline{\omega}_i:\T\rightarrow\R^3$ & Extrinsic tangent vector-field basis\\
  $\omega_i\in\Gamma(T\T)$ & Pulled-back tangent vector-field basis\\
  $\mathbf{m}^\omega\in\R^{3\times3}$ & Element vector mass matrix\\
  $\mathbf{s}^\omega\in\R^{3\times3}$ & Element vector stiffness matrix\\
  $\mathcal{V}\subset\R^3$ & Mesh vertices \\
  $\mathcal{T}\subset[0,|\mathcal{V}|)^3$ & Mesh triangles\\
%  $\mathbf{M}^\psi\in\R^{|\mathcal{V}|\times|\mathcal{V}|}$ & Mesh scalar mass matrix\\
%  $\mathbf{S}^\psi\in\R^{2|\mathcal{V}|\times2|\mathcal{V}|}$ & Mesh scalar stiffness matrix\\
\bottomrule
  \end{tabular}
  \caption
	{
    Notation for per-triangle system construction
    \label{tab:notation}
  }
}
\end{table}
Given a triangle with vertices $v_0,v_1,v_2\in\R^3$, we denote the triangle's normal as:
$$n=\frac{(v_1-v_0)\times(v_2-v_0)}{|(v_1-v_0)\times(v_2-v_0)|}.$$
Computations are performed over the unit right triangle and unit sphere, $\T\subset\R^2$ and $S^2\subset\R^3$:
$$\T=\left\{(s,t)\in[0,1]^2\big|\,s+t\leq1\right\}\quad\hbox{and}\quad S^2=\left\{v\in\R^3\big|\,|v|=1\right\}.$$
The tangent space at $p\in\T$ is equivalent to Euclidean 2-space and the tangent space at $v\in S^2$ is a subset of Euclidean 3-space:
$$T_p\T\simeq\R^2\quad\hbox{and}\quad T_vS^2\subset\R^3.$$
We denote by $\Phi:\T\rightarrow\R^3$ the linear embedding of the triangle:
$$\Phi(s,t) = (1-s-t)\cdot v_0 + s\cdot v_1 + t\cdot v_2$$
and we set $d\Phi|_p$ to be its differential at $p\in\T$: 
$$d\Phi|_p:T_p\T\simeq\R^2\rightarrow T_nS^2\subset\R^3.$$
We set $g|_p$ to be the metric tensor on $T_p\T$ induced by the linear embedding of the unit right triangle:
$$g|_p= d\Phi|_p^\top\cdot d\Phi|_p.$$
We assume that we are given a Gauss Map:
$$N:\T\rightarrow S^2.$$
We denote by $\widetilde{d\Phi}|_p$ the map ``realizing'' the tangent space $T_p\T$ as a subspace of $\R^3$ perpendicular to $N(p)$ by first computing the differential of the embedding $\Phi$ and then applying the Rodrigues rotation taking the normal of the embedded triangle to the normal prescribed by the Gauss Map.
$$\widetilde{d\Phi}|_p\equiv R\big(n,N(p)\big)\circ d\Phi|_p:\R^2\rightarrow\R^3.$$
%(implicitly associating $T_p\T\simeq\R^2$ and $T_{N(p)}S^2\subset\R^3$).
We denote the ``hat'' functions as $\psi_i:\T\rightarrow\R$, with:
$$\psi_0(s,t) = 1-s-t,\quad \psi_1(s,t) = s,\quad\hbox{and}\quad\psi_2(s,t) = t,$$

We note that: (1)~The differential $d\Phi|_p$, and hence the metric $g|_p$ is constant for all $p\in\T$; (2)~The realization $\widetilde{d\Phi}|_p$ is an orthogonal transformation between $T_p\T$ with inner-product $g|_p$ and $T_{N(p)}S^2$ with inner-product obtained by restricting the Euclidean inner-product on $R^3$; (3)~The inverse of $\widetilde{d\Phi}|_p$ is:
\begin{equation}
\label{eq:pull_back}
\widetilde{d\Phi}|_p^{-1}\equiv g^{-1}\cdot d\Phi|_p^\top\cdot R\big(N(p),n\big)
\end{equation}
%(mapping the space perpendicular to $N(p)$ to the space perpendicular to $n$, taking the dot-product with triangle edge directions $(v_1-v_0)$ and $(v_2-v_0)$, and transforming the dot-products into coefficients)
and is defined over all of $\R^3$ by projecting out the component parallel to $N(p)$.

%% file: Sections/04_implementation.tex
\begin{comment}
\begin{figure}[tb]
\begin{center}
    \includegraphics[width=0.3\columnwidth]{Figures/triangle.pdf}
\end{center}
    \caption{The unit-right triangle}
    \label{fig:unit-right-triangle}
\end{figure}
\end{comment}
As our approach is based on defining a basis for function spaces on a mesh, composed of functions that are defined ``piecewise'' per triangle,
%To describe our implementation,
it suffices to consider the case of a single triangle, as the global system can  be constructed using finite-element assembly. To this end, we assume that we are given a triplet of oriented vertices $\{v_k,n_k\}\in\R^3\times S^2$ (with $1\leq k\leq3$), and we extend the per-vertex normals to a Gauss map $N:\T\rightarrow S^2$ by Phong interpolation:
$$N(p) \equiv \frac{\sum_k\psi_k(p)\cdot n_k}{\left|\sum_k\psi_k(p)\cdot n_k\right|}\in S^2.$$
We assume that the Gauss map is nowhere equal to the negative of the triangle's normal, $N(p)\neq-n$, so that $\widetilde{d\Phi}|_p$ is well-defined.

\begin{comment}
In what follows:
\begin{itemize}
\item We denote by $\T$ the unit-right-triangle:
$$\T\equiv\left\{(s,t)\in[0,1]^2\,\big|\,s+t\leq1\right\}.$$
\item We set $D\in\R^{3\times 2}$ to be the matrix of vertex differences:
$$D\equiv\Big(\begin{array}{cc}(v_1-v_0) & (v_2-v_0)\end{array}\Big).$$
\item We set $g\in\R^{2\times 2}\equiv D^\top\cdot D$ to be the symmetric positive definite metric tensor.
\item We set $n\in S^2\equiv (v_1-v_0)\times(v_2-v_0)$ to be triangle's normal.
\item For a function $f:\R^d\rightarrow\R$ we denote by $df\big|_p\in\R^{1\times d}$ the differential of $f$ at $p$ (represented as a row vector).
%\item For (row) vectors $v,w,\in\R^{1\times d}$, we denote by $v\otimes w\equiv v^\top\cdot w\in\R^{d\times d}$ their outer-product.
\item For a matrix $L\in\R^{d\times d}$, we denote by $\hbox{Tr}(L)\in\R$ its trace.
\item For vector $n\in S^d$ and matrix $L\in\R^{d\times d'}$, we denote by $\pi_n(L)\in\R^{d\times d'}$ the matrix obtained by projecting the columns of $L$ onto the space perpendicular to $n$.
\end{itemize}
\end{comment}

\subsection{Scalar Basis}
Using the standard ``hat'' functions gives the mass, $\mathbf{m}^\psi\in\R^{3\times 3}$, and stiffness, $\mathbf{s}^\psi\in\R^{3\times 3}$, matrices:
\begin{align*}
\mathbf{m}_{ij}^\psi&\equiv\int_\T \psi_i(p)\cdot\psi_j(p)\cdot\sqrt{\hbox{det}(g|_p)}\,d\!p\\
\mathbf{s}_{ij}^\psi&\equiv\int_\T \hbox{tr}
\left(
g^{-1}
\cdot
d\psi_i|_p^\top\cdot d\psi_j|_p\right)\cdot\sqrt{\hbox{det}(g|_p)}\,d\!p.
%(d\psi_i|_p\otimes d\psi_j|_p)\right)\cdot\sqrt{|g|}\,dp.
\end{align*}
(The definition of the stiffness is consistent with \eqn{eq:hom_inner_product}, implicitly using the coordinate basis for $\R^3$ and leveraging the fact that, in this basis, the inner-product is represented by the identity matrix.)

\subsection{Vector-Field Basis}
\label{subsec:vector-field-basis}
To define a basis for vector-fields, we proceed as in the approach of Kn\"oppel~\etal \cite{Knoppel:2013:TOG}, (1)~defining an operator taking a tangent vector at a vertex and transporting it into the interior of incident triangles, (2)~selecting a tangent frame at each vertex, and (3)~blending the transported tangent vectors. %As we do this extrinsically, we begin by defining an interpolated normal field in the standard way.

\begin{figure*}[tb]
\begin{center}
    \includegraphics[width=\textwidth]{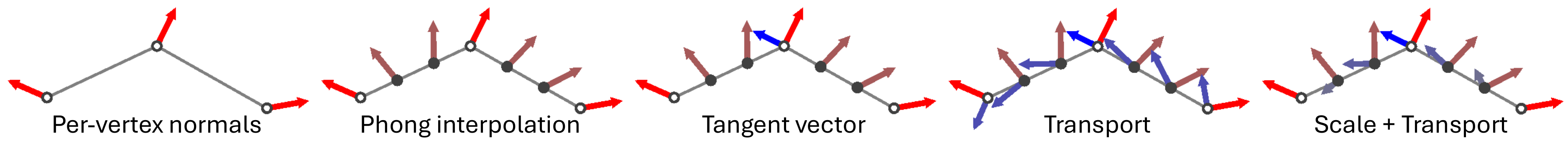}
\end{center}
    \caption{\revision{Constructing a continuous tangent vector-field on a polyline by scaling and transporting a vector prescribed at the center vertex.}}
    \label{fig:diagram}
\end{figure*}

\revision{\fig{fig:diagram} visualizes our approach for a polygonal curve. Starting with per-vertex normals, Phong interpolation assigns normals to points in the interior of the edges. Then, given a tangent vector at the center vertex, the vector is transported using the rotation taking the center vertex's normal to the Phong normals -- giving a vector-field. Finally, a continuous vector-field supported on the two edges is obtained by scaling with the ``hat'' function at the center vertex.}

\paragraph*{Transport}
Given a tangent vector $t_k\in\R^3$ at the $k$-th corner (perpendicular to $n_k$), we define its transport to $p\in\T$ by applying the Rodrigues rotation taking $n_k$ to $N(p)$, to the tangent vector $t_k$. We denote by $R_k:\T\rightarrow\hbox{SO}(3)$ the minimal-angle rotation taking the normal at the $k$-th corner to the interpolated normal at $p$:
$$R_k(p)\equiv R\big(n_k,N(p)\big).$$

\begin{comment}
\subsubsection*{Tangent spaces}
We can associate vectors in the tangent space $T_pT$ with vectors in $\R^3$ by composing the Rodrigues rotation (taking the triangle's normal to the interpolated normal) with the differential of the triangle's embedding:
$$d\Psi|_p\equiv R\big(n,N(p)\big)\cdot D.$$
And, we can define the inverse map as:
$$d\Phi|_p^{-1}\equiv g^{-1}\cdot D^\top\cdot R\big(N(p),n\big).$$

We make several observations about this mapping:
\begin{itemize}
\item The mapping $d\Phi|_p$ from the subspace of $\R^3$ perpendicular to $N(p)$ to $T_p\T$ is an orthogonal transformation (with respect to the standard Euclidean inner-product on $\R^3$ and the inner-product on $T_p\T$ given by $g$).
\item The inverse map $d\Phi^{-1}|_p$ acts on all of $\R^3$, projecting out the component parallel to $N(p)$ before mapping to $T_p\T$.
\item The abuse of notation in denoting the mapping as ``$d\Phi$'' is intentional -- suggesting this is the differential of some mapping $\Phi:\T\rightarrow\R^3$. Indeed, if such a mapping were given, the association between the tangent space $T_p\T$ and $\R^3$ would be defined in terms of its differential. However, in our construction no such mapping is given.
\end{itemize}
\end{comment}

\paragraph*{Framing Vertices}
At the $k$-th corner we select framing vectors, $f_{2k},f_{2k+1}\in\R^3$ perpendicular to the normal $n_k$. For simplicity, we choose the vectors to be orthonormal.% (As in Kn\"oppel~\etal, the choice of per-corner frame does not change the function space.) 

\paragraph*{Blending Transported Tangent Vectors}
We extend the framing vectors, $f_{2k},f_{2k+1}\in\R^3$, to extrinsic vector-fields defined in the interior of the triangle, $\overline{\omega}_i:\T\rightarrow\R^3$, by transporting and blending:
\begin{equation}
\overline{\omega}_{2k+l}(p)\equiv \psi_k(p)\cdot R_k(p) \cdot f_{2k+l}\qquad \hbox{with }l\in\{0,1\}
\label{eq:vf_basis}
\end{equation} 
We note that, by construction, these are perpendicular to the interpolated normals.

\revision
{
To obtain intrinsic vector-fields on the tangent space of the unit right triangle, we pull back using $\widetilde{d\Phi}|_p$:
$$\omega_i|_p\equiv \widetilde{d\Phi}|_p^{-1}\cdot \overline{\omega}_i|_p\in T_p\T.$$

To obtain the vector-fields' covariant derivatives, we recall that that the covariant derivative of an extrinsic vector-field can be computed by differentiating the field and projecting out the normal component. In particular, since $\widetilde{d\Phi}|_p^{-1}$ projects out the normal component, the intrinsic representation of the covariant derivative is obtained by differentiating \eqn{eq:vf_basis} and pulling back:
$$
\nabla\omega_i|_p\equiv \widetilde{d\Phi}|_p^{-1}\circ d\overline{\omega}_i|_p\in\hbox{End}(T_p\T).
$$
}
\begin{comment}
Differentiating \eqn{eq:vf_basis} we get a field giving the change in the 3D vector-field as a function of tangent direction:
$$d\overline{\omega}_i|_p:T_p\T\rightarrow\R^3.$$
We pull these back to get intrinsic vector-fields on the tangent space of the unit right triangle, and their associated covariant derivatives:
\begin{align*}
\omega_i|_p&\equiv \widetilde{d\Phi}|_p^{-1}\cdot \overline{\omega}_i|_p\in T_p\T\\
\nabla\omega_i|_p&\equiv \widetilde{d\Phi}|_p^{-1}\circ d\overline{\omega}_i|_p\in\hbox{End}(T_p\T).
\end{align*}
We recall that as $\widetilde{d\Phi}|_p^{-1}$ projects out the normal component, the covariant derivative is only defined in terms of the component of $d\overline{\omega}_i$ in the tangent space $T_{N(p)}S^2$.
\end{comment}

Using the six functions $\omega_i:\T\rightarrow\R^3$ we compute the mass and stiffness matrices, $\mathbf{m}^\omega,\mathbf{s}^\omega\in\R^{6\times 6}$ with:
\begin{align}
\nonumber
\mathbf{m}_{ij}^\omega&\equiv\int_\T g|_p\big(\omega_i|_p,\omega_j|_p\big)\cdot\sqrt{\hbox{det}(g|_p)}\,d\!p\\
%\mathbf{m}_{ij}^\omega&\equiv\int_\T\langle \overline{\omega}_i|_p,\overline{\omega}_j|_p\rangle\cdot\sqrt{\hbox{det}(g|_p)}\,d\!p\\
\label{eq:vector_stiffness}
\mathbf{s}_{ij}^\omega&\equiv\int_\T \hbox{tr}
\left(
g|_p^{-1}
\cdot (\nabla\omega_i|_p)^*\cdot g|_p \cdot(\nabla\omega_j|_p)\right)
\cdot\sqrt{\hbox{det}(g|_p)}\,d\!p.
\end{align}
\revision{For the mass, the inner-product is computed with respect to the inner-product on the tangent space, $g|_p:T_p\T\rightarrow\T_p^*\T$. For the stiffness, we use the inner-product on the space of endomorphisms $\hbox{End}(T_p\T)$ induced by $g|_p$, reviewed in \eqn{eq:hom_inner_product} of \app{a:endomorphism_decomposition}.}

\begin{comment}
\subsubsection*{Factoring the covariant derivative}
As described in the earlier work of de~Goes~\etal~\cite{deGoes:2014:CGF} the covariant derivative can be factored into two one-dimensional spaces and one two-dimensional space:
\begin{align*}
\nabla\omega
&=
\underbrace
{
\underbrace{
\frac{(\nabla\omega)-g^{-1}\cdot(\nabla\omega)^\top\cdot g}2}_{\hbox{curl (1D)}}
}_{\hbox{anti-symmetric}}\\
&+
\underbrace{
\underbrace{
\frac{\hbox{Tr}(\nabla\omega)\cdot\hbox{Id.}}2}_{\hbox{div (1D) }}
+
\underbrace{
\frac{(\nabla\omega)+g^{-1}\cdot(\nabla\omega)^\top\cdot g}2-\frac{\hbox{Tr}(\nabla\omega)\cdot\hbox{Id.}}2}_{\hbox{traceless (2D)}}
}_{\hbox{symmetric}}
\end{align*}
\end{comment}

\subsection{Integration}
For both the scalar and vector-field basis we compute the coefficients of the mass and stiffness matrices using numeric quadrature~\cite{Taylor:2008:JCAM}. For the mass matrices, this only requires that we can evaluate the basis functions at arbitrary locations in the triangle. For the stiffness matrix, this also requires that we are able to evaluate the differential of the basis functions. This is trivial for the scalar basis (since the differential is constant) and, given the simple expression of the Rodrigues rotation formula, is also straight-forward for the vector-field basis. %In the latter case, we need to take the differential of $R_i:\T\rightarrow\hbox{SO}(3)$. This additionally requires differentiating through the normalization.

%Though the use of quadrature for computing integrals can result in numerical error (e.g. when the integrated function is not a polynomial of an appropriately low degree), it has the advantage of allowing the definition of linear systems in which the right-hand-side (constraints) is not in the span of the function basis (e.g. a texture maps or a neural implicit~\cite{Park:2019:CVPR}).%\MK{Pulled because this is more about scalar functions, which is not our contribution.}

\subsection{Finite Element Assembly}
Given a triangle mesh, $\{\mathcal{V},\mathcal{T}\}$, and given functionality for computing the system matrix associated to a triangle, $L:\mathcal{T}\rightarrow\R^{3K\times3K}$, pseudocode for finite element assembly is summarized in \alg{alg:assembly}: The matrix is initialized (step 1); Then, iterating over all triangles (step 2), the system matrix for the triangle is computed (step 3), and its coefficients are added to the corresponding entries of the system matrix defined over the mesh (steps 4-7).

\RestyleAlgo{boxruled}
\begin{algorithm}
\caption{FiniteElementAssembly}
\label{alg:assembly}
\begin{algorithmic}[1]
\Require 
$L:\mathcal{T}\rightarrow\R^{3K\times3K}$\vspace{-5pt}
\mbox{\hspace{-15pt}\line(1,0){3.08in}}
%\Ensure $y = x^n$
%\State$v\leftarrow0$
\State $\mathbf{L}\leftarrow\mathbf{0}\in\R^{|\mathcal{V}|K\times|\mathcal{V}|K}$
\State\textbf{for} $\tau=\{v_0,v_1,v_2\}\in\mathcal{T}$:
\State\quad$\mathbf{l}\leftarrow L(\tau)$
\State\quad\textbf{for} $m,n\in[0,3)$ and $k,l\in[0,K)$:
\State\quad\quad$I\leftarrow v_m\cdot K+k,\quad J\leftarrow v_n\cdot K+l$
\State\quad\quad$i\leftarrow m\cdot K+k,\quad j\leftarrow n\cdot K+l$
\State\quad\quad$\mathbf{L}_{IJ}\leftarrow\mathbf{L}_{IJ}+\mathbf{l}_{ij}$
\State\textbf{return} $\mathbf{L}$
\end{algorithmic}
\end{algorithm}

For the scalar mass and stiffness matrices, which have one degree of freedom per vertex, we have $K=1$. For the vector-field mass and stiffness matrices we have $K=2$. In what follows, we abuse notation, using $\psi_i$ (resp. $\omega_{2i}$ and $\omega_{2i+1}$), with $i\in[0,|\mathcal{V}|)$, to denote the scalar (resp. vector-field) basis functions on the mesh, rather than at the corners of a single triangle.

%% file: Sections/05_energies.tex
While the previous section described the construction of the stiffness matrix for vector-fields, corresponding to the Dirichlet energy defined by the covariant derivative, a similar approach can be used to define other energies. We also show that our discretization can be used to define the Lie bracket operator.

The following discussion holds for all finite-elements discretization of vector-fields that support evaluation of their derivative. In particular, we discuss constructions of the Killing energy and definitions of the Lie bracket derived from the finite-elements proposed by Stein~\etal \cite{Stein:2020:CGF} (ignoring discontinuities along edges). While in principle a similar approach can be used with the finite-elements of Kn\"oppel~\etal \cite{Knoppel:2013:TOG}, the lack of a closed-form expression for the basis functions makes this challenging in practice. This issue is addressed in the later work of Liu~\etal~\cite{Liu:TOG:2016}.

\subsection{Energies}
\label{ss:energies}
Using the decomposition from \sec{ss:linear_algebra}, we can factor the covariant derivative into orthogonal components consisting of scalar multiples of the identity, trace-free (self-adjoint) endomorphisms, and anti-self-adjoint endomorphisms. This, in turn, allows us to replace the covariant derivative in \eqn{eq:vector_stiffness} with an individual component -- allowing us to define classical energies used in vector-field processing such as:
\begin{align*}
\hbox{Connection}\,\,\,\,&\longleftrightarrow\,\,\,\,\scalar{T_p\T}\oplus\noscalar{T_p\T}\oplus\asym{T_p\T}\\
\hbox{Holom. / Hodge}\,\,\,\,&\longleftrightarrow\,\,\,\,\scalar{T_p\T}\oplus\asym{T_p\T}\\
\hbox{Anti-holom.}\,\,\,\,&\longleftrightarrow\,\,\,\,\noscalar{T_p\T}\\
\hbox{Killing}\,\,\,\,&\longleftrightarrow\,\,\,\,\scalar{T_p\T}\oplus\noscalar{T_p\T}\\
\hbox{Divergence}\,\,\,\,&\longleftrightarrow\,\,\,\,\scalar{T_p\T}\\
\hbox{Curl}\,\,\,\,&\longleftrightarrow\,\,\,\,\asym{T_p\T}\\
\end{align*}
We note that the spaces $\scalar{T_p\T}$ and $\asym{T_p\T}$ are one-dimensional, while the space $\noscalar{T_p\T}$ is two-dimensional.

\begin{comment}

\begin{figure}
    \centering
    \includegraphics[width=.8\linewidth]{Figures/projection.pdf}
    \caption{A random vector-field (top left) decomposed into its divergence, curl, and anti-holomorphic components.}
    \label{fig:projection}
\end{figure}

An immediate consequence of this is that it is straight-forward to extract the different components of a vector-field. For example, given a vector-field represented by the coefficients $\mathbf{z}\in\R^{2|\mathcal{V}|}$, the ``divergence-full'' component is obtained by solving:
$$\mathbf{S}^\omega\cdot\mathbf{z}_{\hbox{\tiny div}} = \mathbf{S}^\omega_{\hbox{\tiny div}}\cdot\mathbf{z}$$
with $\mathbf{z}_{\hbox{\tiny div}}\in\R^{2|\mathcal{V}|}$ the coefficients of the projected vector-field and $\mathbf{S}^\omega,\mathbf{S}^\omega_{\hbox{\tiny div}}\in\R^{2|\mathcal{V}|\times2|\mathcal{V}|}$ the stiffness matrices integrating the square-norm of the covariant derivative and the square-norm of the component of the covariant derivative in $\scalar{T_p\T}$, respectively. As an example, \fig{fig:projection} shows a random vector-field (top left) and its decomposition into its $\scalar{T_p\T}$ (top right), $\asym{T_p\T}$ (bottom left), and $\noscalar{T_p\T}$ (bottom right) components.
\end{comment}

\subsection{Lie Bracket}
Our approach provides an expression for the covariant derivative that can be evaluated pointwise. In addition to being amenable to quadrature-based integration, it also enables the pointwise evaluation of the Lie bracket of two vector-fields. Concretely, given tangent vector-fields $X$ and $Y$, we evaluate the Lie bracket at a point $p\in\T$ by taking the difference of the derivative of $X$ along $Y$ and the derivative of $Y$ along $X$:
$$[X,Y](p) = \nabla X|_p\cdot Y(p)- \nabla Y|_p\cdot X(p).$$
\revision{We note that this is only an approximation of the Lie Bracket, since this relation is exact only when $\nabla$ is the Levi-Civita Connection.}
\subsubsection*{Global fitting}
As our space of tangent vector-fields is not closed under the Lie bracket, the bracket of vector-fields $X$ and $Y$ cannot (in general) be expressed as a linear combination of our tangent vector-field basis. However, it is straightforward to compute the projection of the bracket $[X,Y]$ onto our space of vector-fields.

Given tangent vector-fields $X$ and $Y$, we start by computing the ``weak representation'' of their Lie bracket, $\mathbf{b}\in\R^{2|\mathcal{V}|}$, by computing the integral of the dot-product of the bracket with the tangent vector-field basis functions (which only requires that the bracket be pointwise evaluatable):
$$\mathbf{b}_i\equiv\int
%\hbox{tr}\left(\omega_i^\top(p)\cdot g|_p\cdot[X,Y](p)\right)
g|_p\big(\omega_i(p),[X,Y](p)\big)
\cdot\sqrt{\hbox{det}(g|_p)}\,d\!p.$$
Then, we obtain the coefficients of the projection of the bracket onto the space of tangent vector-fields, $\mathbf{z}\in\R^{2|\mathcal{V}|}$, by solving:
\begin{equation}
\mathbf{M}^\omega\cdot\mathbf{z} = \mathbf{b}.
\label{eq:bracket_fit}
\end{equation}

%\paragraph{Note} A similar approach can be used to compute the projection of the covariant derivative of one vector-field with respect to another, onto $\mathcal{W}$.

%% file: Sections/06_topology.tex
\subsection{Parallel transport and Gaussian curvature}
%Discussing the induced parallel transport and geodesy of straight segments
As our vector-fields are continuous and we interpolate normals, \revision{their derivatives} induce notions of \revision{per-triangle piecewise} continuous parallel transport and Gaussian curvature. Consider a single triangle $\tau\in \mathcal{T}$ with
%vertices $\{v_i,v_j,v_k\}$ and respective
normals $\{n_1,n_2,n_3\}$. Without loss of generality, consider the total rotation from and back to the first vertex:
$$\exp(\kappa_\tau\cdot n_1) = R(n_3,n_1)\cdot R(n_2,n_3)\cdot R(n_1,n_2).$$
Then, $\kappa_\tau$ is the \emph{holonomy} of the triangle with the prescribed normals. 
Note that since our tangent bundle is decoupled from the geometry of the mesh, the total sum of curvature might have a different Euler characteristic than the mesh. \revision{For instance, it is 0 when the vertex normals are constant. By definition, it is however always a multiple of $2\pi$. % that is computed from summing up the holonomies.
For our ``natural'' vertex normals, we expect it to match the Euler characteristic of the geometry when the mesh is decently-sampled, but we do not make this assumption.}

\begin{figure}
    \includegraphics[width=0.24\textwidth]{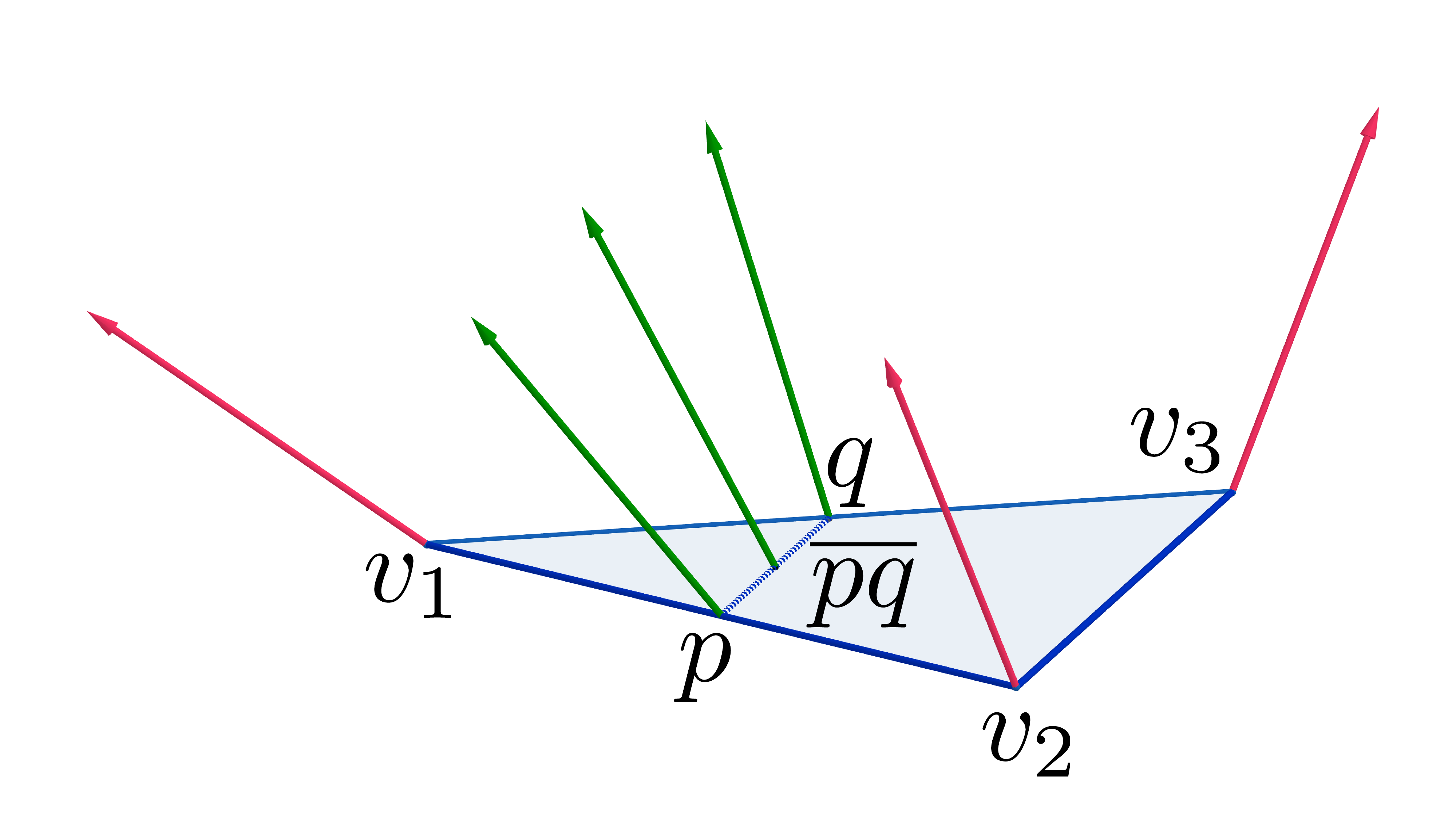}
    \includegraphics[width=0.24\textwidth]{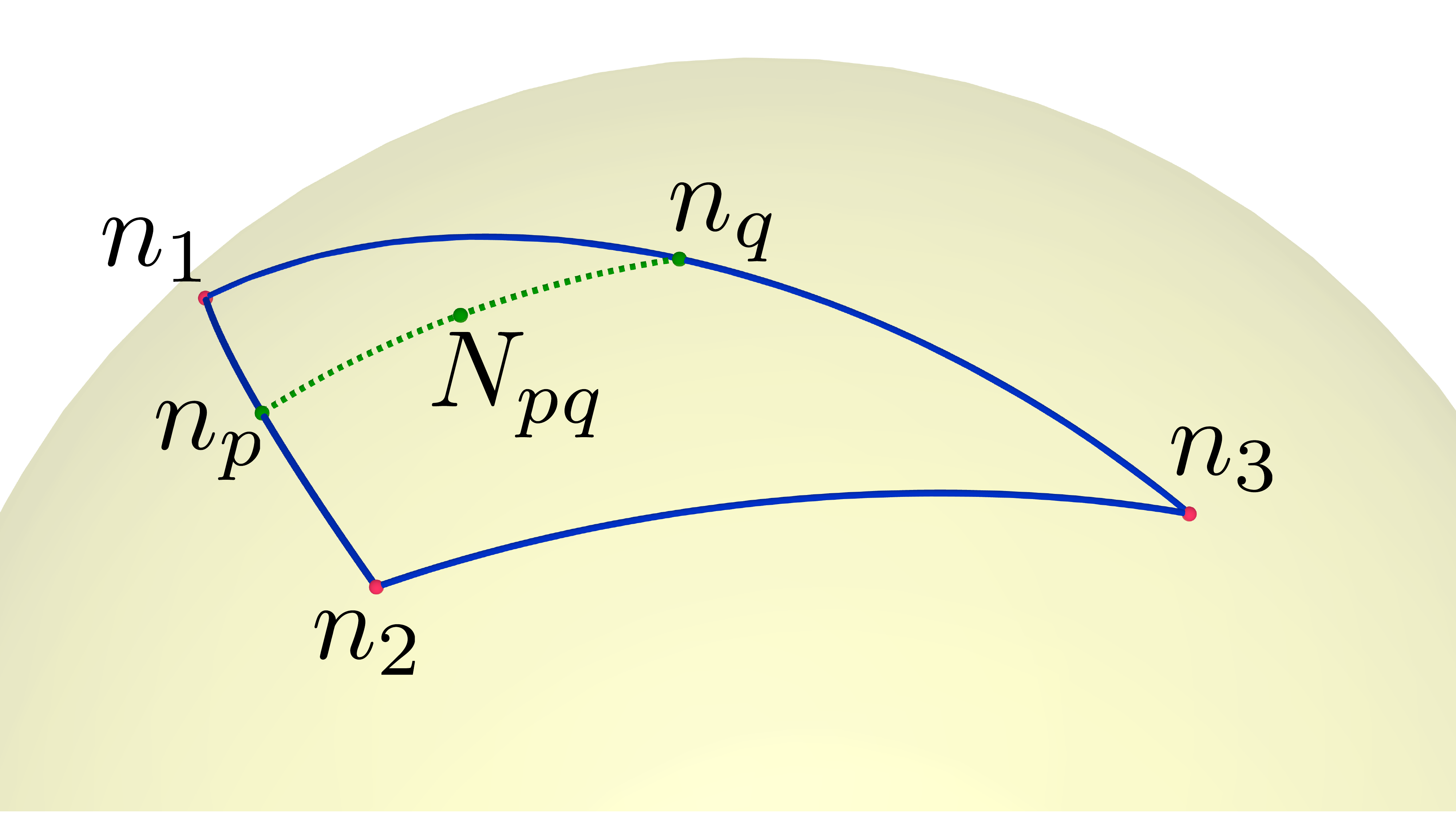}
    \caption{\revision{Interpolated normals (green) along a segment on the triangle map onto a great circle on the Gauss map.}}
    \label{fig:great-circle}
\end{figure}

The holonomy $\kappa_\tau$ can be computed directly as the angle defect of the geodesic triangle in the Gauss map~\cite{do_carmo_differential_2016}. We use the equivalent solid angle formula for the induced area:
%\MK{Should we pull the structure-preservation comment, given that it assumes we have good normals to start with? Also, do we rely on this property, e.g. for stating Equation 6?}
\begin{equation}
\kappa_\tau  =  2 \operatorname{atan2}\!\left(
  n_1 \cdot (n_2 \times n_3),\;
  1
  + (n_1 \cdot n_2)
  + (n_2 \cdot n_3)
  + (n_3 \cdot n_1)
\right)
\label{eq:normal-holonomy}
\end{equation}
%\MK{Do we need this definition of holomony, given that we have a workable definition above?}
\begin{remark}
Given a triangle $\tau\in\T$, the normals along any straight segment within the triangle, $\overline{pq}\subset\tau$, all lie on a great circle  \revision{(\fig{fig:great-circle})}.
%To see this, consider the segment with endpoints $p$ and $q$, having interpolated normals $n_p$ and $n_q$ respectively. Since the normals are interpolated by linear blend followed by pointwise normalization, every normal on the line $(1-s)p+sq,\ \forall s\in[0,1]$ is in the plane spanned by $n_p$ and $n_q$, and thus forming a great circle in the Gauss map.
\end{remark}
\begin{proof}
Let $\alpha_k:\T\rightarrow\R$ be the barycentric coordinate functions. For any point $\overline{pq}(s)=(1-s)\cdot p+s\cdot q$ on the segment, the interpolated (un-normalized) normal is:
$$
\widetilde{N}\big(\overline{pq}(s)\big)
=\sum_{k=1}^3\alpha_k(\overline{pq}(s))\cdot n_k
=(1-s)\cdot\tilde{n}_p+s\cdot\tilde{n}_q
%\left(\sum_{k=1}^3\alpha_k(p)\cdot n_k\right)+s\cdot\left(\sum_{k=1}^3\alpha_k(q)\cdot n_q\right).
$$
where $\tilde{n}_p$ and $\tilde{n}_q$ are the (un-normalized) vectors obtained by linearly interpolating the normals from the corner of the triangle to $p$ and $q$ respectively. Thus, the normals along the segment $\overline{pq}$ all lie in the plane spanned by $\tilde{n}_p$ and $\tilde{n}_q$, and hence (since the span is independent of normalization) on a great circle. \hfill\end{proof}

A direct implication is that the parallel transport between $p$ and $q$ is simply $R(n_p,n_q)$, where $n_p$ and $n_q$ are the unit normals at $p$ and $q$ obtained by interpolating from the corners and normalizing. \revision{More formally, the Gauss map of the interpolated normals is a part of a sphere, where segments map to geodesics (parts of great circles). As a consequence, the parallel transport along $pq$ is the pullback of parallel transport along the sphere, as defined in Sec.~\ref{subsec:vector-field-basis}.}

This allows us to measure holonomy on any sub-triangle simply by using \eqn{eq:normal-holonomy} on the interpolated normals at the sub-triangle's vertices, making the algorithm of the next section simple to implement.

%\revision{Our parallel transport is the pullback of parallel transport along the sphere. Parallel transport along a segment of a great circle on the sphere is simply R(np, nq), so our parallel transport along a straight line segment is also R(np, nq)?}

\subsection{Singularities}
\label{subsec:singularities}
\begin{figure}
    \centering
\includegraphics[width=1.0\linewidth]{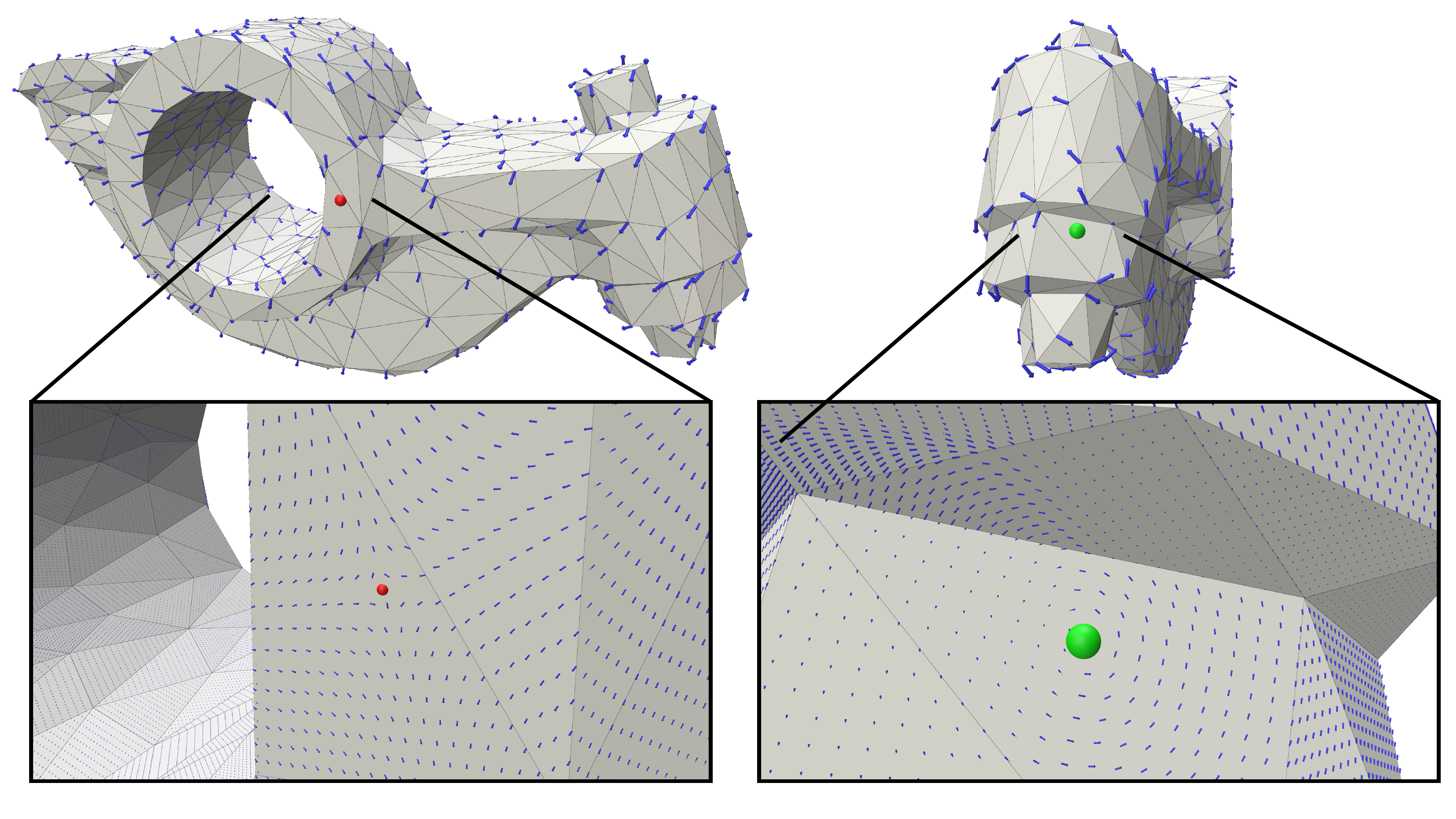}
    \caption{We detect positive (green) and negative (red) singularities in the interior of faces.}
    \label{fig:singularities}
\end{figure}

Given a vector field $\omega$, for every triangle $\tau\in\mathcal{T}$~\cite{Crane:2010:TCD}, the index of the field over the triangle is:
$$
I_\tau^\omega =\frac{1}{2\pi}\left(\theta_\tau^\omega - \kappa_\tau\right),
%I_\tau^\omega =\frac{1}{2\pi}\left(\theta_{12}^\omega+\theta_{23}^\omega+\theta_{31}^\omega - \kappa_\tau\right),
$$
where $\theta_\tau^\omega$ is the sum of the rotation angles of the field deviating from parallel transport, taken over the edges of $\tau$.
%They can be easily computed by assigning arbitrary local 2D bases at $i$ and $j$ (without loss of generality).
Since the field is continuous, singularities can appear anywhere (as in~\cite{Boksebeld2022}). To find them, we rely on the fact that every segment within the triangle is a great circle on the Gauss map, so that the angle sum $\theta^\omega$ and holonomy $\kappa$ can be computed in closed form for any subtriangle. Thus, we can find the barycentric coordinates $\{\alpha_1,\alpha_2,\alpha_3\}$ of a singularity
using bisection. We show examples in Fig.~\ref{fig:singularities}. The definition of the total index of a triangle is exact; \revision{we further assume that the pointwise singularities are as those of a standard PL field: either a single point per triangle, a full line, or the entire triangle (we omit the last two cases). We leave a concrete (dis)proof for future work.}

%\MK{Does this statement "every segment within the triangle" need to be highlighted, e.g. as a Lemma?}
%, described in \alg{alg:singularity}.  Essentially, we are looking for the coordinate $\alpha_k$ such that the constant barycentric segment it defines in $\tau$ crosses the singularity.

To find where coordinate $\alpha_k$ crosses the singularity, \alg{alg:singularity} proceeds as follows:
Assuming there is a singularity inside the triangle (step 1), the search interval is initialized (steps 2 and 3), and bisection is iterated until the interval length is within a prescribed tolerance (steps 4-9). Within each iteration, the mid-point is identified (step 5), the index of the triangle having $k$ as one of its vertices and the positions $\alpha_k$ along the edges incident to vertex $k$ as the other two vertices is computed (step 6), and the interval is refined to be the sub-interval with the singularity (steps 7 and 8). We use a tolerance of $\varepsilon=10^{-7}$. We search using $k=1$ and with $k=2$, thereby locating the singularity within the triangle.%\MK{Are we assuming a single singularity per triangle?}

%by the following bisection algorithm on two barycentric coordinates
%(defined on $\alpha_i$ without loss of generality):\MB{use an algorithm env}\MK{Should make notation consistent with the start of section 4}
\begin{comment}
\begin{enumerate}
    \item if the index $I_{ijk}=0$, the triangle is regular (no singularities), terminate.
    \item $\alpha_{i,\text{start}} := 0,\ \alpha_{i, \text{end}}:=1$,
    \item $\alpha_{i,\text{mid}} = \frac{1}{2}\left(\alpha_{i,\text{start}}+\alpha_{i,\text{end}}\right)$,
    \item Compute the index $I_{i,\text{mid}}$ of the subtriangle created by vertex $i$ and the edge points at coordinates $(\alpha_{i,\text{mid}},  1 - \alpha_{i,\text{mid}}, 0)$ (on edge $ij$) and $(\alpha_{i,\text{mid}},  0, 1 - \alpha_{i,\text{mid}})$ (on edge $ki$),
    \item if $I_{i,\text{mid}} \neq 0$, set $\alpha_{i,\text{start}} = \alpha_{i,\text{mid}}$, otherwise set $\alpha_{i,\text{end}} = \alpha_{i,\text{mid}}$.
    \item if $\alpha_{i,\text{end}} - \alpha_{i,\text{start}}<\varepsilon$ terminate.\MB{give the tolerance a name}
    \item Jump to (3).
\end{enumerate}
\end{comment}

%\begin{comment}
\RestyleAlgo{boxruled}
\begin{algorithm}
\caption{SingularityDetection}
\label{alg:singularity}
\begin{algorithmic}[1]
\Require 
$k\in\{1,2,3\}$,\quad $\varepsilon>0$\vspace{-5pt}
\mbox{\hspace{-15pt}\line(1,0){3.08in}}
\State\textbf{if} $I_\tau^\omega=0$: \textbf{return} false
\State$\alpha_\text{start}\leftarrow0$
\State$\alpha_\text{end}\leftarrow1$
\State\textbf{while} $\alpha_\text{end}-\alpha_\text{start}\geq\varepsilon$:
\State\quad$\alpha_\text{mid}\leftarrow\frac{1}{2}\left(\alpha_\text{start}+\alpha_\text{end}\right)$
\State\quad$I\leftarrow\text{SubTriangleIndex}(k,\alpha_{\text{mid}})$\ \ (Eq.\ref{eq:normal-holonomy})
\State\quad\textbf{if} $I\neq0$:\quad
$\alpha_{\text{end}}\leftarrow\alpha_\text{mid}$
\State\quad\textbf{else}:\,\,\,\quad\quad
$\alpha_{\text{start}}\leftarrow\alpha_\text{mid}$
\State\textbf{return} \{true, $\alpha_\text{mid}$\}
\end{algorithmic}
\end{algorithm}
%\end{comment}

%We use $\varepsilon = 10^{-7}$. The barycentric coordinate of the singularity is then reported as $\alpha_{i,\text{mid}}$. We do the same for $\alpha_j$, which pinpoints the exact singularity location.

\subsection{N-fields}

\begin{figure}
    \centering
\includegraphics[width=1.0\linewidth]
{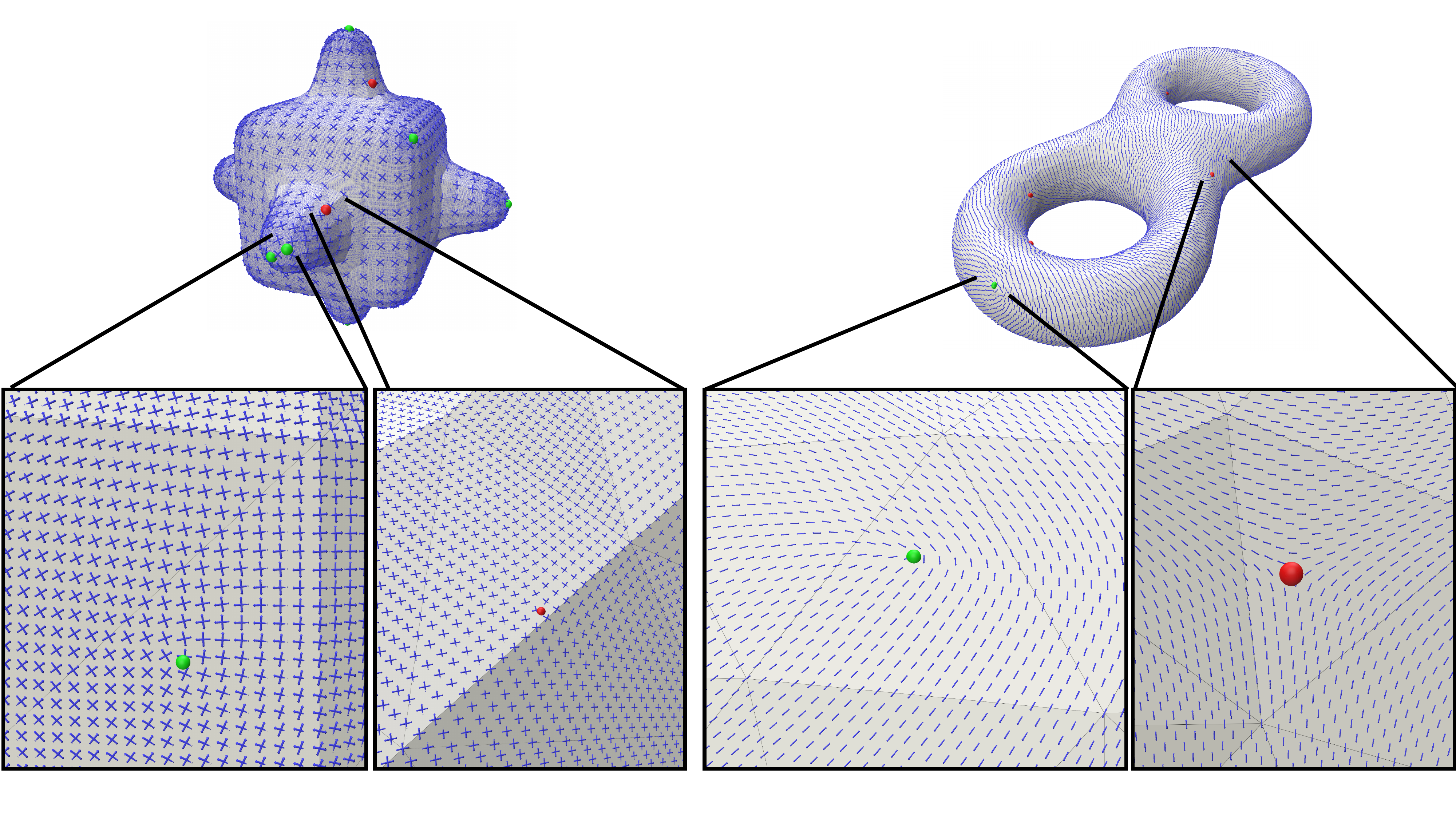}
    \caption{Our framework allows the interpolation of $N$-fields, and locating their continuous singularities. Left: $N=4$, right: $N=2$.}
    \label{fig:N-singularities}
\end{figure}

Given a triangle $\tau\in\mathcal{T}$, we represent $N$-RoSy fields~\cite{Vaxman2016} using the orthonormal framing vectors $f_{2k}$ and $f_{2k+1}$ at the $k$-th vertex \revision{(Sec.~\ref{subsec:vector-field-basis})}, allowing us to define the power representation $X = \omega^N$ of an $N$-RoSy vector in each vertex's tangent plane. At each internal point $p$, we choose an arbitrary orthonormal frame, defining the connection $r_{k,p} \in \mathbb{C}$ as the representation of the frame $\{R(n_k,N(p))\cdot f_{2k+l}\}_l$ with respect to the frame of $p$. Then, we get:
$$
X_{p} = \sum_{k\in\tau} X_k\cdot \left(r_{k,p}\right)^N,
$$
and the $N$-RoSy field at $p$ as the roots $\sqrt[N]{X_p}$. Singularities are discovered as in \sec{subsec:singularities}, with two modifications: (1) we compute $\theta^X$ as the angle sum defined by $X$, rather than $\theta^\omega$, and (2) we use $N\cdot\kappa_\tau$ as the holonomy. The obtained integer indices are interpreted as integer multiples of $1/N$. This is standard practice in defining $N$-RoSy singularities~\cite{Diamanti2014}. We show examples in Fig.~\ref{fig:N-singularities}. We note that PolyVectors can be defined similarly.%, and reserve this demonstration for future work.

%% file: Sections/07_evaluation.tex
We evaluate our discretization of the space of tangent vector-fields in four ways: (1)~We compare to existing approaches in standard interpolation applications;  (2)~We validate that our discretization preserves some basic invariants from the continuous theory; (3)~We consider the spectral decomposition of the the connection Laplacian and associated operators; and (4)~We demonstrate applications in computing the Lie bracket of vector-fields.

In our implementation, we use 3-point quadrature to integrate functions over a triangle. With the exception of the unit sphere we rescale all models to have unit area.
We assign normals to vertices by computing the approximate limit surface under Loop subdivision~\cite{Loop:1987}. Concretely, for each vertex we compute the subdivision stencil, raised to the $10$-th power, and apply it to the vertex and its one-ring neighbors. We then set the vertex's normal to the area-weighted average of the normals of the subdivided triangles.
%We assign normals by first computing the area-weighted average of the normals of incident faces and then performing a short time-step (average edge length divided by 200) of heat diffusion~\cite{Prada:2016:TOG}. And, 

With the exception of the spectral analysis on the sphere, we use the same geometry for all methods. In practice, this means that the method of Stein~\etal, which assigns two degrees of freedom to each edge, has roughly three times as many degrees of freedom as the methods that assign two degrees of freedom to a vertex.

When ground-truth is available, we measure the error between the estimated solution, $x$ and the ground-truth, $x^*$ as the ratio:
$$E(x,x^*)\equiv\sqrt{\frac{\|x-x^*\|^2}{\|x\|^2+\|x^*\|^2}}$$
where, for a function/field $f$ defined on the surface, ``$\|f\|^2$'' denotes the integrated square-norm over the mesh.

We visualize vector-fields using anisotropic diffusion \cite{Diewald:TVCG:2000} in which initially randomly distributed noise at the vertices is smoothed in the direction of the flow-field. Similar to line integral convolution \cite{Cabral:CGIT:1993}, the visualization captures the direction of the vector-field but not the sign. To address this, we overlay the visualization with stream-lines, shaded from black (start) to white (end). 
%Additionally, by having the diffusion kernel grow inversely proportionally with vector-field magnitude, one obtains a visualization in which flow-lines become more blurred where the vectors are smaller -- thereby giving a sense of size as well as direction. As with line integral convolution, the visualization does not reveal the sign of the vectors.

%We begin by considering several applications from geometry processing before proceeding to an empirical evaluation.

To support follow-on research we provide an implementation in C++, using a functional-programming paradigm, that makes it straight-forward to represent fields over the mesh (\href{https://github.com/mkazhdan/PhongRodriguesVF/}{https://github.com/mkazhdan/PhongRodriguesVF/}). The implementation only requires that a user provide a functor whose input is a position on the mesh (defined by the triangle index and barycentric coordinates) and whose output is the field's value at that point. (These fields can themselves be constructed by evaluating vector-fields and their covariant derivatives to construct the field's values, as in the case of computing the Lie bracket.) Using quadrature, the implementation integrates the fields, performing the finite-element assembly required to construct system matrices and vectors for targeted vector-field-processing applications.

We compare our approach to that of Kn\"oppel~\etal \cite{Knoppel:2013:TOG}, Stein~\etal \cite{Stein:2020:CGF}, and Sharp~\etal \cite{Sharp:2019:TOG}, with implementations provided by the authors, \cite{geometrycentral,fieldgen} We also compare to the discretization obtained using the Whitney basis \cite{Whitney:GIT:1957}. We could not compare to Liu~\etal \cite{Liu:TOG:2016} as a public implementation is not available.

\subsection{Example Applications}
\label{ss:example_applications}

\subsubsection*{Sparse Vector-Field Interpolation}
\begin{figure}[tb]
\begin{center}
    \includegraphics[width=\columnwidth]{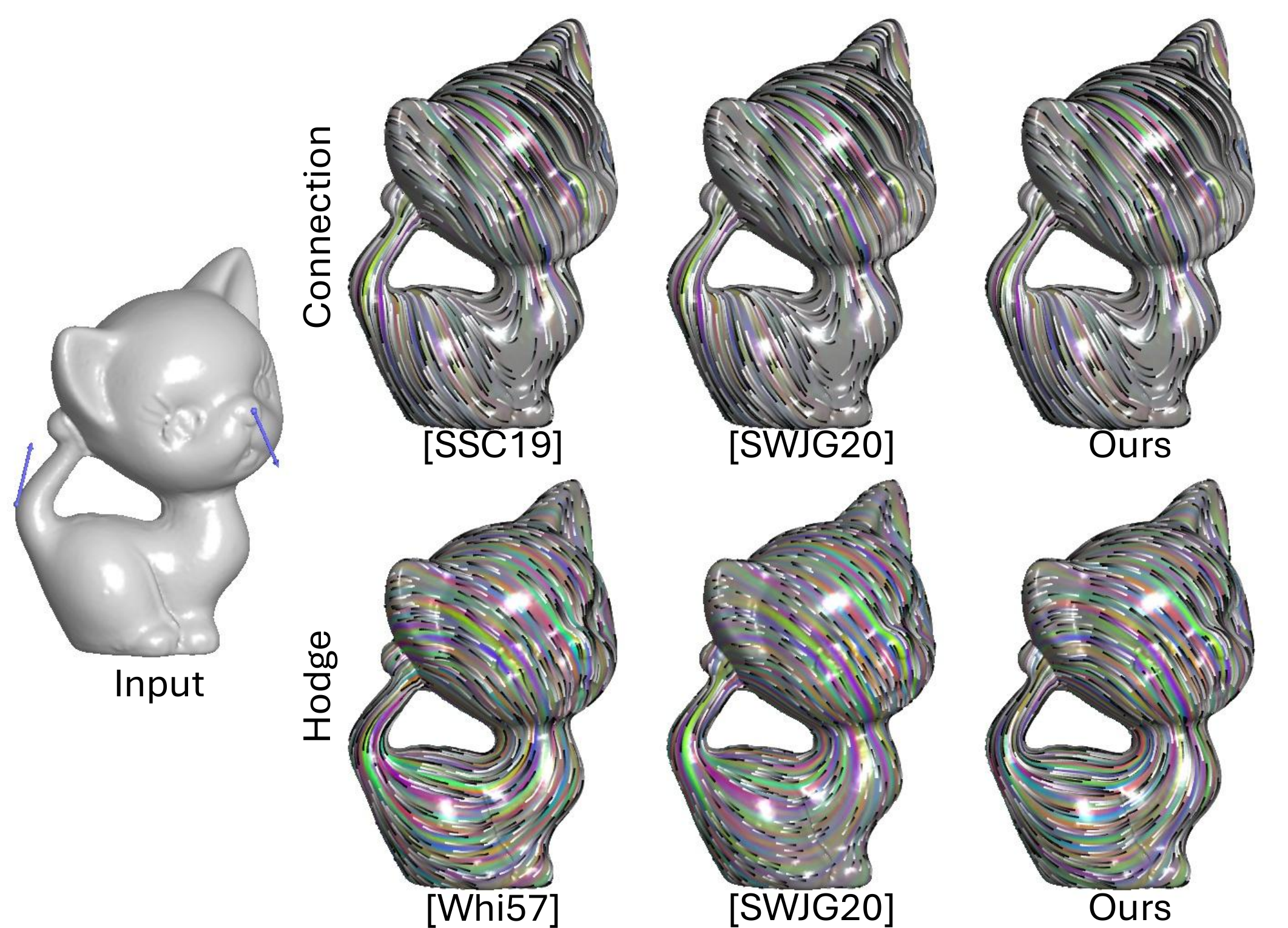}
\end{center}
    \caption{Visualization of smooth interpolation of sparse vectors (draw in blue) using smoothness energies defined by the Connection (top) and Hodge (bottom) Laplacians.
    \label{fig:interpolation}}
\end{figure}
Given a sparse subset of vertices $\mathcal{W}\subset\mathcal{V}$ and target tangent vectors at those vertices $\{t_w\}_{w\in\mathcal{W}}$, the stiffness matrix can be used to solve for the as-smooth-as-possible interpolating vector-field:
$$\mathbf{x} = \argmin_{\mathbf{y}\in\R^{3|\mathcal{V}|}}\left(\mathbf{y}^\top\cdot\mathbf{S}\cdot\mathbf{y}\right),\quad\hbox{s.t.}\quad\mathbf{y}_w=t_w,\,\forall w\in\mathcal{W}$$
with $\mathbf{S}$ a stiffness matrix.

As an example, \fig{fig:interpolation} shows a model with two tangent vector constraints (blue) and the computed vector-fields obtained with $\mathbf{S}$ defined by the connection (top) and Hodge (bottom) energies. As the model has genus one, there are two harmonic vector-fields and Hodge interpolation gives the vector-field circulating around the hole. In contrast, the connection energy, which is non-singular, penalizes the vector-field wherever it is non-zero, resulting in an interpolant that gets smaller away from the constraints.

The top row compares our approach to that of Sharp~\etal (second column) and Stein~\etal (third column). As the figure shows, all three approaches produce a similar sparse interpolant. %(Variations in the visualization are due to the fact that line integral convolution randomly seeds colors at vertices, so small variations in vector-field results in differently colored lines.)

Since the approach of Stein~\etal provides a pointwise evaluation of the covariant derivative, we can similarly define a Hodge energy by considering the holomorphic component of the covariant derivative (ignoring the unbounded discontinuity at edges). Interpolation results obtained using that approach are shown in the bottom row. We also show results obtained using the DEC formulation of the Hodge Laplacian associated with the Whitney 1-form basis \cite{Whitney:GIT:1957,Bossavit:CE:1998,Desbrun:2006:SigCourse}. We find that our results are more consistent with those obtained using the Whitney basis. For the approach of Stein~\etal it was necessary to add a connection energy regularizer, as otherwise the system was not well-conditioned. This is consistent with the authors' incorporation of a connection regularizer to remove spurious minima in the Killing energy. 

For a more detailed empirical discussion, please see \app{a:sparse_interpolation}.

\subsubsection*{Vector Heat}
\begin{figure}[tb]
\begin{center}
    \includegraphics[width=\columnwidth]{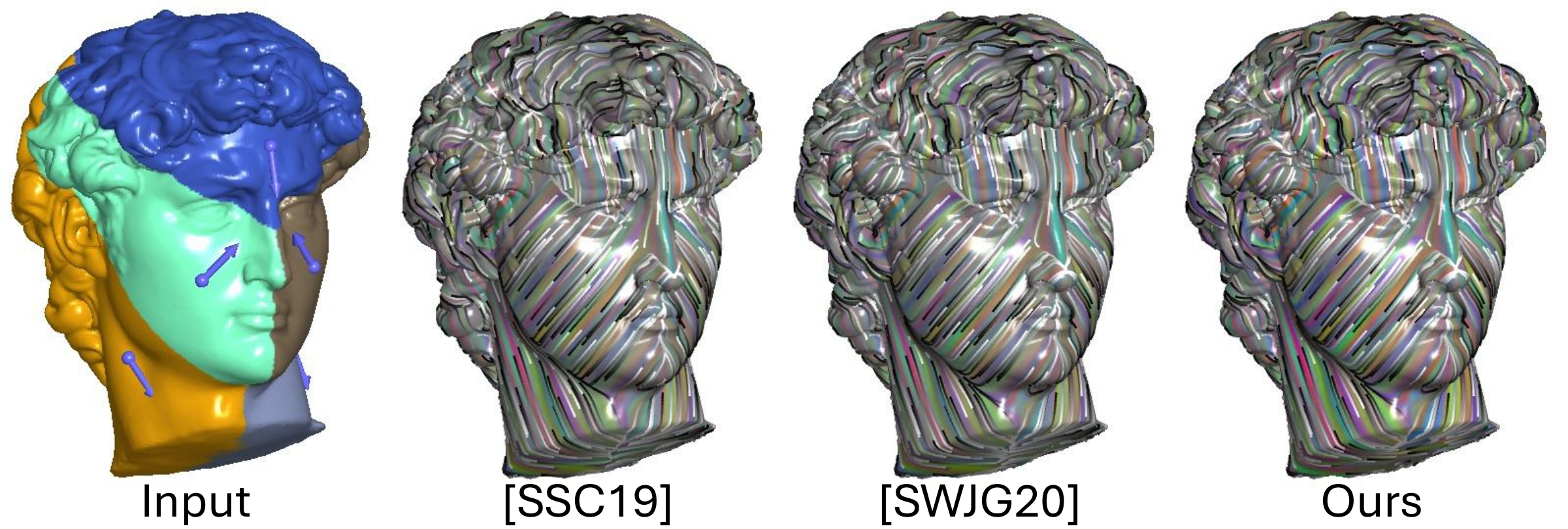}
\end{center}
    \caption{Input vector-field constraints, with the associated Voronoi partition of the geometry (left), and the results of applying the Vector Heat method using different discretizations of the connection Laplacian.}
    \label{fig:heat}
\end{figure}
We also consider the Vector Heat method for parallel transport of vectors \cite{Sharp:2019:TOG}. The method is implemented using both vector heat diffusion (using the connection Laplacian) and scalar heat diffusion (using the co-tangent Laplacian). The result is a vector-field on the surface whose value within a Voronoi region of a constraint point is the parallel transport of the constraint vector along the shortest path geodesic.

\fig{fig:heat} shows the results for five constraint vectors placed on the surface of the David model. The visualization shows the initial constraint vectors and associated Voronoi regions (left) as well as the output vector-field. As with sparse vector-field interpolation, we compare our results (right) to those of Sharp~\etal (center left) and Stein~\etal (center right). Empirically, we find almost no difference between the approaches. This is consistent with the earlier observation that, when using the connection stiffness for sparse vector-field interpolation, the approaches give similar results.

\subsection{Rotation Invariance}
\label{ss:rotation_invariance}
%\subsubsection*{In-plane Rotation}
As with the discretizations of Kn\"oppel~\etal and Stein~\etal, our function space is closed under pointwise rotation by a fixed angle in the tangent plane. In particular, letting $\mathbf{M},\mathbf{T},\overline{\mathbf{T}},\overline{\mathbf{A}},\mathbf{J}\in\R^{2|\mathcal{V}|\times2|\mathcal{V}|}$ be the matrices corresponding to the:
$$
\begin{array}{c@{\,\,\,}l}
\mathbf{M} & \hbox{mass,}\\
\mathbf{T} & \hbox{stiffness of component $\scalar{T_p\T}$,}\\\overline{\mathbf{T}} & \hbox{stiffness of component $\noscalar{T_p\T}$,}\\
\overline{\mathbf{A}} & \hbox{stiffness of component $\asym{T_p\T}$, and}\\
\mathbf{J} & \hbox{$90^\circ$ counter-clockwise rotation around each vertex's normal,}
\end{array}
$$
we validate that the mass matrix and trace-less component are both invariant under counter-clockwise rotation by $90^\circ$ and that the scalar-multiple of the identity and anti-self-adjoint components are related to each other by $90^\circ$ counter-clockwise rotation. 

For the models in \fig{fig:hodge.0}, we show the relative errors in \tab{tab:splitting}. The table validates the expected properties of the system matrices under counter-clockwise rotation by $90^\circ$. The tiny relative errors for the mass matrix are expected, as the vector-field representation is extrinsic and the Euclidean inner-product is invariant under rotation. For the other energies, the relative error is also small, with the largest errors for the ``bunny'' and ``icosa'' models. We believe this is due to the quality of the triangulations (both models were obtained by applying Marching Cubes \cite{Lorensen:1987:SIGGRAPH} to an implicit representation). In particular, considering the distribution of triangle aspect ratios in \fig{fig:aspect_ratios}, we find that the error is strongly correlated with triangulation quality. (As $\mathbf{J}\cdot\mathbf{J}=\mathbf{J}^\top\cdot\mathbf{J}^\top=-\mathbf{Id}$, it follows that $\mathbf{T}=\mathbf{J}^\top\cdot\overline{\mathbf{A}}\cdot\mathbf{J}$ if and only if $\overline{\mathbf{A}}=\mathbf{J}^\top\cdot\mathbf{T}\cdot\mathbf{J}$. This was empirically confirmed by noting that the relative errors are nearly identical.)

\begin{table}
\small
\center{
  \begin{tabular}{l|ccc}
  Model
  & 
  $
\frac
{
\|\mathbf{M}-\mathbf{J}^\top\cdot\mathbf{M}\cdot\mathbf{J}\|_F
}
{
\|\mathbf{M}+\mathbf{J}^\top\cdot\mathbf{M}\cdot\mathbf{J}\|_F
}
  $
  &
  $
\frac
{
\|\overline{\mathbf{T}}-\mathbf{J}^\top\cdot\overline{\mathbf{T}}\cdot\mathbf{J}\|_F
}
{
\|\overline{\mathbf{T}}+\mathbf{J}^\top\cdot\overline{\mathbf{T}}\cdot\mathbf{J}\|_F
}
  $
  & 
  $
\frac
{
\|\mathbf{T}-\mathbf{J}^\top\cdot\overline{\mathbf{A}}\cdot\mathbf{J}\|_F
}
{
\|\mathbf{T}+\mathbf{J}^\top\cdot\overline{\mathbf{A}}\cdot\mathbf{J}\|_F
}$
  \\
\midrule
Bunny &
$1.5\times10^{-16}$ & $1.3\times10^{-10}$ & $1.0\times10^{-10}$ \\
Eight &
$1.5\times10^{-16}$ & $2.2\times10^{-16}$ & $2.3\times10^{-16}$ \\
Elephant &
$1.4\times10^{-16}$ & $1.6\times10^{-15}$ & $1.3\times10^{-15}$ \\
Genus-6 &
$1.4\times10^{-16}$ & $4.6\times10^{-16}$ & $3.7\times10^{-16}$ \\
Icosa &
$1.4\times10^{-16}$ & $4.2\times10^{-10}$ & $3.1\times10^{-10}$ \\
  \end{tabular}
  \caption
	{
    Invariance under $90^\circ$-rotation, computed using the ratio of the Frobenius norm of the difference to the Frobenius norm of the sum.
    \label{tab:splitting}
  }
}
\end{table}

\subsection{Spectral Analysis}
\label{ss:spectral}
\subsubsection*{Spectral Vector-Fields}

Using the energies from \sec{ss:energies}, we solve the generalized eigenproblem $\mathbf{S}\cdot\mathbf{x} = \lambda\cdot\mathbf{M}\cdot\mathbf{x}$ where $\mathbf{M}$ is the mass matrix and $\mathbf{S}$ is one of the connection, Hodge, and Killing stiffness matrices.

\begin{figure}[tb]
\begin{center}
    \includegraphics[width=\columnwidth]{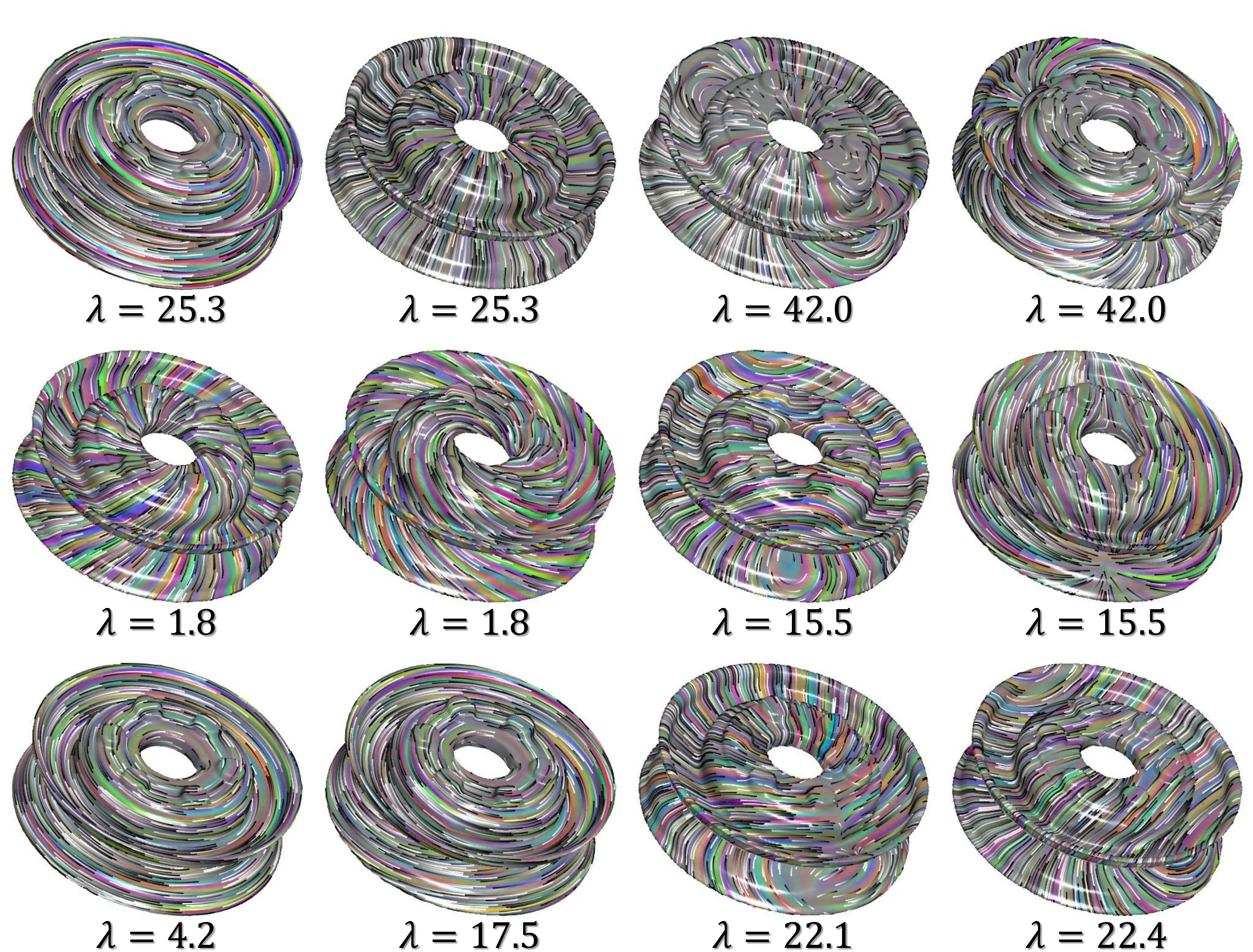}
\end{center}
    \caption{Visualization of The four smallest eigenvectors of the Connection (top), Hodge (middle), and Killing (bottom) energies.}
    \label{fig:spectra}
\end{figure}

\fig{fig:spectra} shows the Pulley model and the smallest four eigenvectors of each energy, along with the associated eigenvalues. We note that:
\begin{itemize}
\item Since the connection and Hodge energies use equal parts of $\scalar{T_p\T}$ and $\asym{T_p\T}$, their eigenvectors come in pairs, related by $90^\circ$ counter-clockwise rotation (see above).
\item Since the model has genus one, the two smallest eigenvectors of the Hodge Laplacian closely match the harmonic vector-fields and their associated eigenvalues are close to zero.
\item Since the model is nearly a surface of revolution, the smallest eigenvector of the Killing energy corresponds to the flow around the axis of revolution, and its associated eigenvalue is noticeably smaller than that of the next eigenvector. (Though the second eigenvector also corresponds to a vector-field circulating about the axis of rotational symmetry, the directions of flow are reversed on the top and bottom halves of the pulley.)
\end{itemize}

For visualization, we further decompose the eigenspaces of the connection and Hodge energies. To this end, we set $\mathbf{M}\in\R^{2|\mathcal{V}|\times2|\mathcal{V}|}$ to be the mass matrix and $\mathbf{T}\in\R^{2|\mathcal{V}|\times2|\mathcal{V}|}$ to be the stiffness matrix associated with component $\scalar{T_p\T}$ -- defining the quadratic energy giving the integrated squared divergence of a vector-field. Then, for a particular eigenspace, $X$, of dimension $2k$, letting $\mathbf{X}^{2|\mathcal{V}|\times2k}$ be the matrix whose columns are the eigenvectors, we define matrices $\mathbf{m},\mathbf{t}\in\R^{2k\times2k}$ with:
$$\mathbf{m}=\mathbf{X}^\top\cdot\mathbf{M}\cdot\mathbf{X}\qquad\hbox{and}\qquad\mathbf{t}=\mathbf{X}^\top\cdot\mathbf{T}\cdot\mathbf{X}.$$
Solving the generalized eigenvalue problem:
$$\mathbf{t}\cdot\mathbf{x}=\lambda\cdot\mathbf{m}\cdot\mathbf{x}$$
we obtain an orthonormal basis for the eigenspace $X$ graded by divergence energy. In particular, letting $\mathbf{x}_1,\ldots\mathbf{x}_{2k}$ be the computed eigenvectors, we order the eigenvectors as $k$ pairs, with the first vector minimizing squared divergence and the second its $90^\circ$ counter-clockwise rotation:
$$\{\mathbf{X}\cdot\mathbf{x}_1,\mathbf{J}\cdot\mathbf{X}\cdot\mathbf{x}_1,\ldots,\mathbf{X}\cdot\mathbf{x}_k,\mathbf{J}\cdot\mathbf{X}\cdot\mathbf{x}_k\}.$$

\subsubsection*{Comparison to Stein et al.}
%In general, any finite-elements discretization supporting pointwise evaluation of vector-fields and their derivatives can be used to define a Killing energy, (defined as the integrated square norm of the symmetric component of the covariant derivative). In particular, we evaluated the energy defined by Crouzeix-Raviart elements of Stein~\etal \cite{Stein:2020:CGF}. 

In principle, the formulation of Stein \etal \cite{Stein:2020:CGF} can also be used to define a Killing energy. However, as discussed in that work, the direct definition of such an energy exhibits spurious high-frequency minima (likely due to aliasing arising from the discontinuous nature of the basis). This limitation can be mitigated by adding a connection energy regularizer to the Killing energy. However, that introduces the additional challenge of tuning the regularization weight. Please see \app{a:bracket} for additional discussion. 

\subsubsection*{Spectral Analysis on the Sphere}
To study the spectral decomposition of the connection Laplacian defined over the sphere, we generate ten random triangulations of the unit sphere in two ways: For the first, we randomly sample points on the sphere and triangulate the points by computing the convex hull. For the second, we randomly sample points on the ellipsoid with semi-axis lengths $(1,4,1)$, triangulate the points by computing the convex hull, and rescale the points to have unit-norm. We then compute the mass and connection Laplacian matrices, $\mathbf{M}$ and $\mathbf{S}$ and solve the generalized eigenvalue problem:
$$\mathbf{S}\cdot\mathbf{x} = \lambda\cdot\mathbf{M}\cdot\mathbf{x}$$
to obtain the first 240 eigenvalues.

As the $n$-th eigen-space of the connection Laplacian is $(4n+2)$-dimensional, with associated eigenvalue $n(n+1)-1$, we evaluate the connection Laplacian by measuring the difference of the estimated eigenvalues from the ground-truth.

\begin{figure}[tb]
\begin{center}
    \includegraphics[width=\columnwidth]{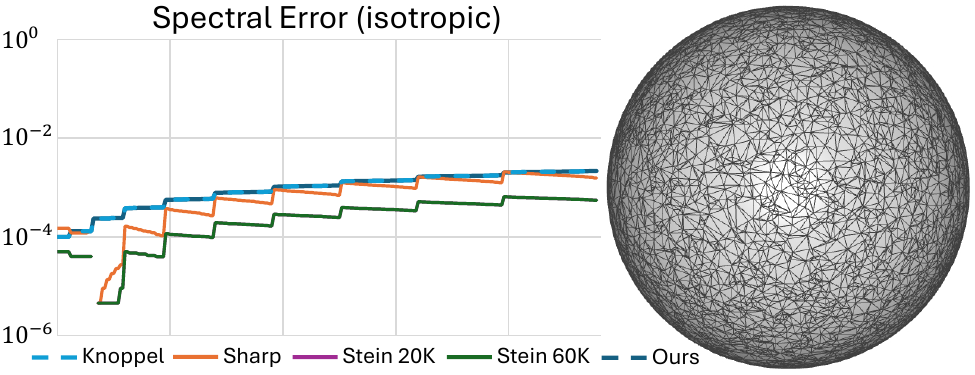}
    \includegraphics[width=\columnwidth]{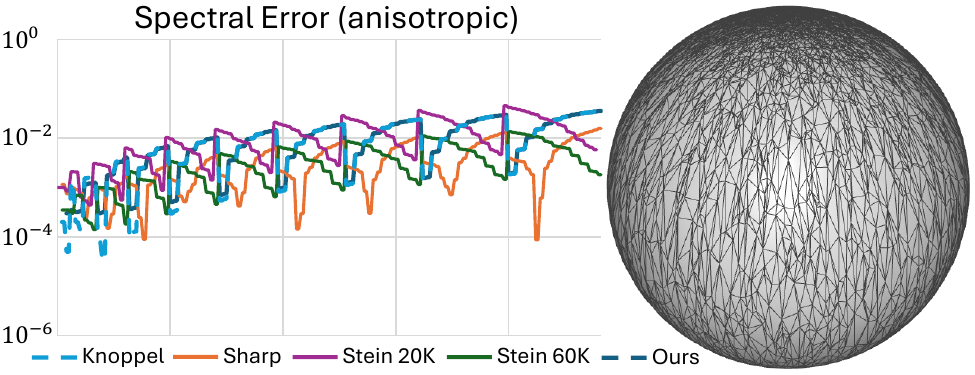}
\end{center}
    \caption{The differences between the analytic eigenvalues and the estimated ones, averaged over ten random tessellations of the unit sphere, as a function of eigenvalue index. The top chart plots results when the sphere is sampled isotropically. The bottom plots results for the more challenging case of anisotropic sampling. (The associated eigenvalues are $\{1,5,11,19,29,41,55,71,89,109\}$.) Representative tessellations of the sphere are shown on the right.}
    \label{fig:spectrum-error}
\end{figure}

\fig{fig:spectrum-error} compares our results to those of Kn\"oppel~\etal \cite{Knoppel:2013:TOG}, Stein~\etal \cite{Stein:2020:CGF}, and Sharp~\etal \cite{Sharp:2019:TOG}. For the visualization, we plot the ratio $|ev_c-ev_{gt}|/|ev_c+ev_{gt}|$, with $ev_c$ the computed eigenvalue and $ev_{gt}$ the ground-truth. Since the mass matrix of Sharp~\etal is not obtained using a finite-elements approach, it may not be positive definite when the mesh is not Delaunay, and we apply intrinsic Delaunay triangulation \cite{Fisher:2006:SIGCourse} before computing their system matrices. For all methods, we discretize the connection Laplacian over a sphere sampled with $60K$ points. For the method of Stein~\etal, we also show results for a sphere sampled with $20K$ points, so that the number of degrees of freedom match (two degrees of freedom per edge for Stein~\etal, versus two degrees of freedom per vertex for the other methods).

For both the isotropically (top) and anisotropically (bottom) sampled sphere, our results are indistinguishable from those of Kn\"oppel~\etal and are comparable to those of the method of Stein~\etal when using $20K$ samples. (Using a higher resolution of $60K$ samples, the discretization of Stein~\etal gives better results, as expected.) While also comparable to the method of Sharp~\etal for the anisotropically sampled sphere, we find that the spectrum of Sharp~\etal's Laplacian is closer to that of the ground-truth in the case of anisotropic sampling. We believe that this is due to the use of the intrinsic Delaunay triangulation, which has the effect of undoing the anisotropic triangulation.

%In computing the spectrum using the methods of Sharp~\etal and Stein~\etal, we used the lumped (i.e. diagonal) mass matrix, whereas the unlumped matrix was used for Kn\"oppel~\etal and ours. As has been observed by others, we found that using the lumped matrix results in an under-estimate of the eigenvalues while using the unlumped matrix results in over-estimate (particularly for the isotropic sampling). This is visualized in \fig{fig:spectrum-error} by the tendency of the error of our discretization and that of Kn\"oppel~\etal to increase within an eigenblock, while the error of Sharp~\etal and Stein~\etal tends to decrease. Replacing the unlumped matrix with the lumped matrix (which is reasonable under the assumption that the tangent space is framed using an orthonormal basis) similarly resulted in an under-estimate of the eigenvalues, but did not noticeably affect the magnitude of the errors. \MK{Do we need a citation for "As has been observed by others"? Though interesting, should we pull the paragraph?}

\subsubsection*{Hodge Laplacian}
\begin{figure*}[tb]
\begin{center}
    \includegraphics[width=2.1\columnwidth]{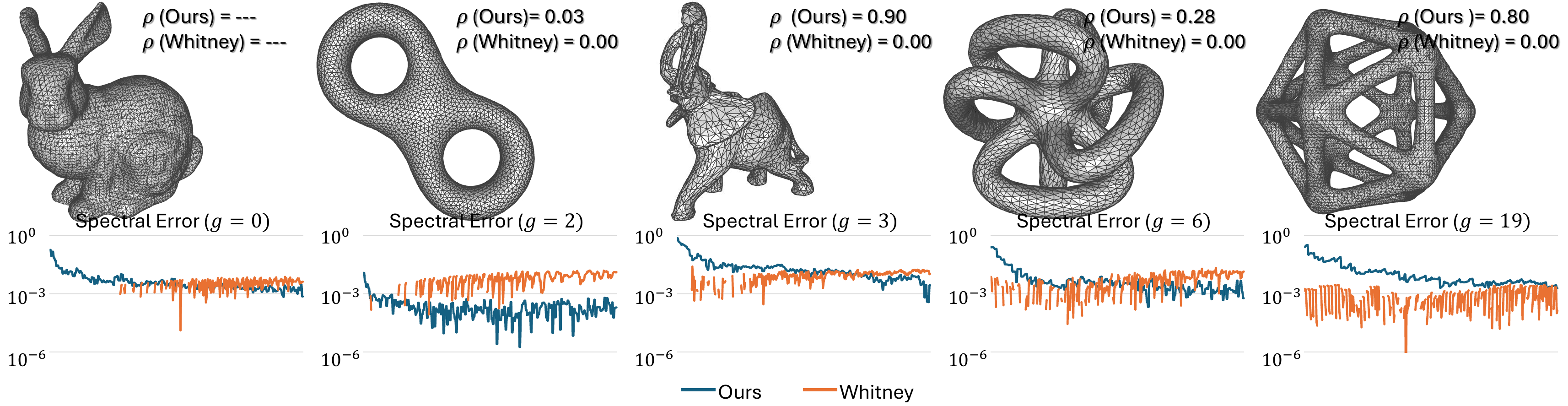}
\end{center}
    \caption{The relative difference between the eigenvalue of the Hodge Laplacian and the co-tangent Laplacian, computed using the holomorphic part of the connection Laplacian and the discretization using the Whitney basis. The plot also gives the genus, $g$, of the model and the ratio, $\rho$, of eigenvalues $2g$ and $2g+1$.}
    \label{fig:hodge.0}
\end{figure*}

Considering just the holomorphic component of the covariant derivative (i.e. the sum of the projections onto $\scalar{T_\mathbf{p}\T}$ and $\asym{T_\mathbf{p}\T}$ we obtain a quadratic Hodge energy.

We measure the similarity between the spectra of our holomorphic energy and the spectra of the co-tangent Laplacian. We compare with the spectra obtained using the Hodge Laplacian computed using the Whitney basis. %For both methods, we use the lumped mass matrix for 0-forms, since the definition of the Hodge energy using the DEC formulation requires inverting this matrix.

\fig{fig:hodge.0} shows results for a number of shapes, with different genuses. We plot the ratio $|ev_h-ev_c|/|ev_h+ev_c|$ where $ev_h$ is the eigenvalue of the Hodge Laplacian and $ev_c$ is the associated eigenvalue of the co-tangent Laplacian, for the first 200 eigenvalues. Concretely, for a genus $g$ surface, we compare  eigenvalues $2(g+i)$ and $2(g+i)+1$ of the Hodge Laplacian with eigenvalue $i+1$ of the co-tangent Laplacian, accounting for the $2\cdot g$ harmonic vector-fields and the constant scalar field in the kernels of the Laplacians.

In addition to comparing the spectra, the plots also give the ratio, $\rho$, of the $2g$-th and $(2g+1)$-st eigenvalues of the Hodge Laplacian. Since we expect to have $2g$ harmonic vector-fields, this ratio measures the extent to which the Hodge Laplacian fails to identify the harmonic vector-fields.

As the Whitney 1-forms are consistent with the structure-preserving DEC discretization, they exactly identify the harmonics ($\rho=0$). Furthermore, since the differentials of the eigenvectors of the co-tangent Laplacian are themselves eigenvectors of the Hodge Laplacian, every eigenvalue of the co-tangent Laplacian will necessarily be an eigenvalue of the Hodge Laplacian obtained using the DEC formulation. However, since the Whitney 1-forms are not closed under rotation by $90^\circ$ degrees, the eigenvalues obtained using the DEC discretization do not come in pairs and it is \textit{not} the case that the $(i+1)$-st eigenvalue of the co-tangent Laplacian will appear as either the $(2g+2i)$-th or $(2g+2i+1)$-st eigenvalue of the Hodge Laplacian discretized using DEC. Thus, while the DEC discretizations tend to give errors where every other one is initially zero, this ceases to hold at higher frequencies.

More generally, we found that at lower frequencies, the spectrum of the Whitney basis discretizations does a better job of matching the spectrum of the co-tangent Laplacian, while at higher frequencies our discretization does better.

To understand the behavior of our discretizations under refinement, we \revision{repeated} the experiment comparing the spectrum of the Hodge Laplacian to the spectrum of the co-tangent Laplacian, performing multiple passes of Loop subdivision~\cite{Loop:1987}.
%\fig{fig:hodge.elephant} shows the results for the genus-3 elephant model.
We \revision{found} that both our and the Whitney energies \revision{improved} with refinement, and the extent of improvement \revision{appeared} to be comparable. We also \revision{found} that the trend of the Whitney basis outperforming our discretization at lower frequencies persists. Unfortunately, the ability of our discretization to distinguish the harmonic vector-fields \revision{did} not improve consistently under refinement.

\begin{comment}
\begin{figure}[tb]
\begin{center}
    \includegraphics[width=.8\columnwidth]{Figures/hodge.elephant.pdf}
\end{center}
    \caption{Comparison of relative eigenvalue differences computed for the ``Elephant'' model, with $\iota\in\{0,\ldots,4\}$ passes of Loop subdivision. For our discretization, the ratios of eigenvalues $2g$ and $2g+1$ are, respectively, $\{0.89,0.69,0.85,0.45,0.70\}$.}
    \label{fig:hodge.elephant}
\end{figure}
\end{comment}

\begin{figure}[tb]
\begin{center}
    \includegraphics[width=.8\columnwidth]{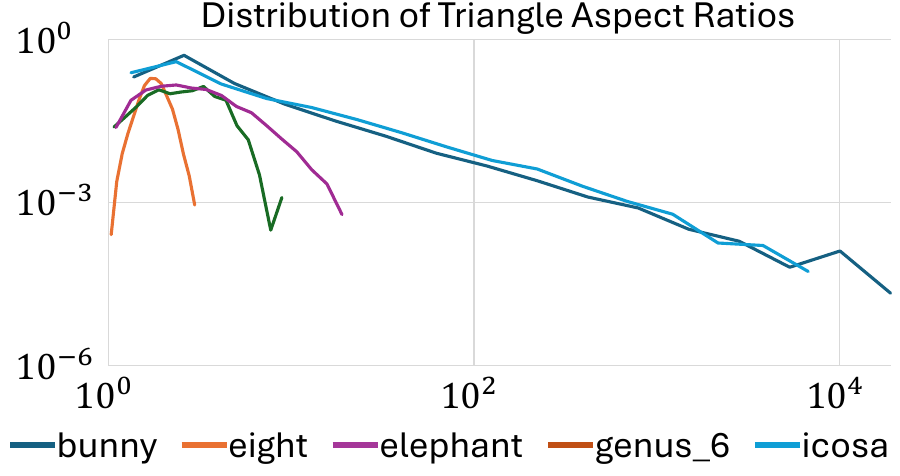}
\end{center}
    \caption{Distribution of triangle aspect ratios for the models visualized in \fig{fig:hodge.0}.}
    \label{fig:aspect_ratios}
\end{figure}

\begin{comment}
\paragraph*{Second Fundamental Form} We further validate our implementation by recalling that the second fundamental form and ambient covariant derivative are related by:
$$\hbox{II}\left(V,W\right)=\langle dV\cdot W,N\rangle.$$
Setting $X_1$ and $X_2$ to be the coordinate-fields
$$X_1(p)=\left(\begin{array}{c}1\\0\end{array}\right)\in T_p\T
\quad\hbox{and}\quad
X_2(p)=\left(\begin{array}{c}0\\1\end{array}\right)\in T_p\T$$
we obtain a pointwise estimate of the second fundamental form, by realizing the vector-fields $X_1$ and $X_2$ as vector-fields in $\R^3$. Concretely, we get:
$$\hbox{II}_{ij}=\left\langle d\left(\widetilde{d\Phi}\cdot X_i\right)\cdot X_j,N\right\rangle,$$
which can be evaluated analytically given the closed-form expression for $\widetilde{d\Phi}^{-1}$ (\eqn{eq:pull_back}).

\begin{wrapfigure}{r}{0.15\textwidth}
    \hspace{-.38in}
%    \centering
    \includegraphics[width=0.18\textwidth]{Figures/bunny.png}
\end{wrapfigure}
As an alternate expression, we can directly differentiate the Gauss map to get:
$$\hbox{II}=d\Phi^{\top}\cdot dN.$$
Comparing the two definitions over the ``Stanford Bunny'' model described above, we get a relative difference of $1.5\times10^{-14}$.
\end{comment}

\subsection{Lie Bracket}
\label{ss:lie_bracket}
To evaluate our discretization of the Lie bracket, we compare to three approaches. The first two define the bracket indirectly, representing derivations in terms of functional maps \cite{Azencot:CGF:2013,Azencot:TOG:2015}. The third is obtained analogously to ours, differentiating the finite-elements representation of Stein~\etal \cite{Stein:2020:CGF} to obtain the covariant derivative and evaluating the expression $[X,Y](p)=\nabla X|_p\cdot Y(p) - \nabla Y|_p\cdot X(p)$. For both finite-elements discretizations the obtained bracket $[X,Y]$ is not in the span of the basis and we obtain the least-squares fit by solving the linear system in \eqn{eq:bracket_fit} using the respective mass matrices. For our approach this requires approximating integrals using quadrature. For Stein \etal's discretization the bracket is piecewise linear, and integration over the interior of triangles is computed in closed-form.

We evaluate the discretizations of the Lie bracket on two tessellations of the sphere. For the first we recursively subdivide and refine the vertices of an icosahedron. For the second we randomly sample points on the sphere and compute the convex hull. Both tessellations contain roughly 10K vertices.

As input we use vector-fields defined in the ambient 3D space. We start by generating random vector-fields $\tilde{X},\tilde{Y}:\R^3\rightarrow\R^3$, whose Fourier coefficients are band-limited complex exponentials with uniformly randomly assigned Fourier coefficients. (Antipodal frequencies are assigned conjugate values so that the coordinate functions are real-valued.) Setting $\pi$ to be the map projecting vector-fields onto the tangent space of the unit sphere:
$$[\pi(Z)](p)\equiv Z(p)-p\cdot\langle Z(p),p\rangle,$$
we compute the projected vector-fields $X=\pi(\tilde{X})$ and $Y=\pi(\tilde{Y})$.
and obtain their Lie bracket as:
$$[X,Y]\equiv \pi\left(\nabla X \cdot Y - \nabla Y \cdot X\right)$$
with $\nabla$ the Euclidean gradient applied to the coordinates of the vector field. Because $\tilde{X}$ and $\tilde{Y}$ are linear combinations of trigonometric functions, and because the projection $\pi$ has a simple expression, the input vector-fields $X$ and $Y$, and the ground-truth bracket $[X,Y]$ can be evaluated in closed-form.

\begin{table}
\small
\center{
  \begin{tabular}{@{}c@{\,\,\,\,}r|cccc@{}}
   &
  $b$
  & \cite{Azencot:CGF:2013}
  & \cite{Azencot:TOG:2015}
  & \cite{Stein:2020:CGF}
  & Ours
  \\
\midrule
\multirow{4}{*}{icosa}
& 2 &
$3.5\times10^{-3}$ & $9.5\times10^{-4}$ & $\mathbf{1.0\times10^{-4}}$ & $7.7\times10^{-4}$ \\
& 5 &
$8.0\times10^{-3}$ & $1.5\times10^{-3}$ & $\mathbf{3.7\times10^{-4}}$ & $2.4\times10^{-3}$ \\
& 10 &
$2.6\times10^{-2}$ & $3.7\times10^{-3}$ & $\mathbf{7.3\times10^{-4}}$ & $8.9\times10^{-3}$ \\
& 20 &
$9.6\times10^{-2}$ & $1.1\times10^{-1}$ & $\mathbf{1.4\times10^{-3}}$ & $3.5\times10^{-2}$ \\
\midrule
\multirow{4}{*}{CH}
& 2 &
$1.7\times10^{-1}$ & $3.0\times10^{-1}$ & $\mathbf{9.6\times10^{-3}}$ & $1.3\times10^{-2}$ \\
& 5 &
$1.2\times10^{-1}$ & $3.1\times10^{-1}$ & $\mathbf{1.5\times10^{-2}}$ & $1.6\times10^{-2}$ \\
& 10 &
$1.3\times10^{-1}$ & $4.0\times10^{-1}$ & $\mathbf{2.2\times10^{-2}}$ & $3.4\times10^{-2}$ \\
& 20 &
$2.0\times10^{-1}$ & $4.6\times10^{-1}$ & $\mathbf{2.9\times10^{-2}}$ & $9.6\times10^{-2}$ \\
  \end{tabular}
  \caption
	{
    Relative error in estimating the Lie bracket for of two vector-fields for different band-widths, $b$, and different discretizations of the sphere, ``icosa'' vs. ``CH''.
    \label{tab:bracket}
  }
}
\end{table}

\tab{tab:bracket} gives the relative errors for the estimated bracket, computed for band-widths $b\in\{2,5,10,20\}$, using the two tessellations of the sphere. (For \cite{Azencot:TOG:2015} we computed the brackets using spectral dimensions varying from $10$ to $1000$ and give the smallest error.)

As expected, the table shows that discretization accuracy deteriorates as vector-fields become higher frequency and as the tessellation becomes less uniform. In addition, we see that while all methods perform comparably on the subdivided icosahedron, the operator-based approaches are less robust in the case of non-uniform sampling. Interestingly, despite the discontinuity of their representation, we find that the finite-elements discretization of Stein \etal provides the best results. (It is possible that some of the performance is due to the fact that Stein \etal's basis has approximately $3\times$ the numbers of degrees of freedom.)

\begin{figure*}[tb]
\begin{center}
    \includegraphics[width=0.9\linewidth]{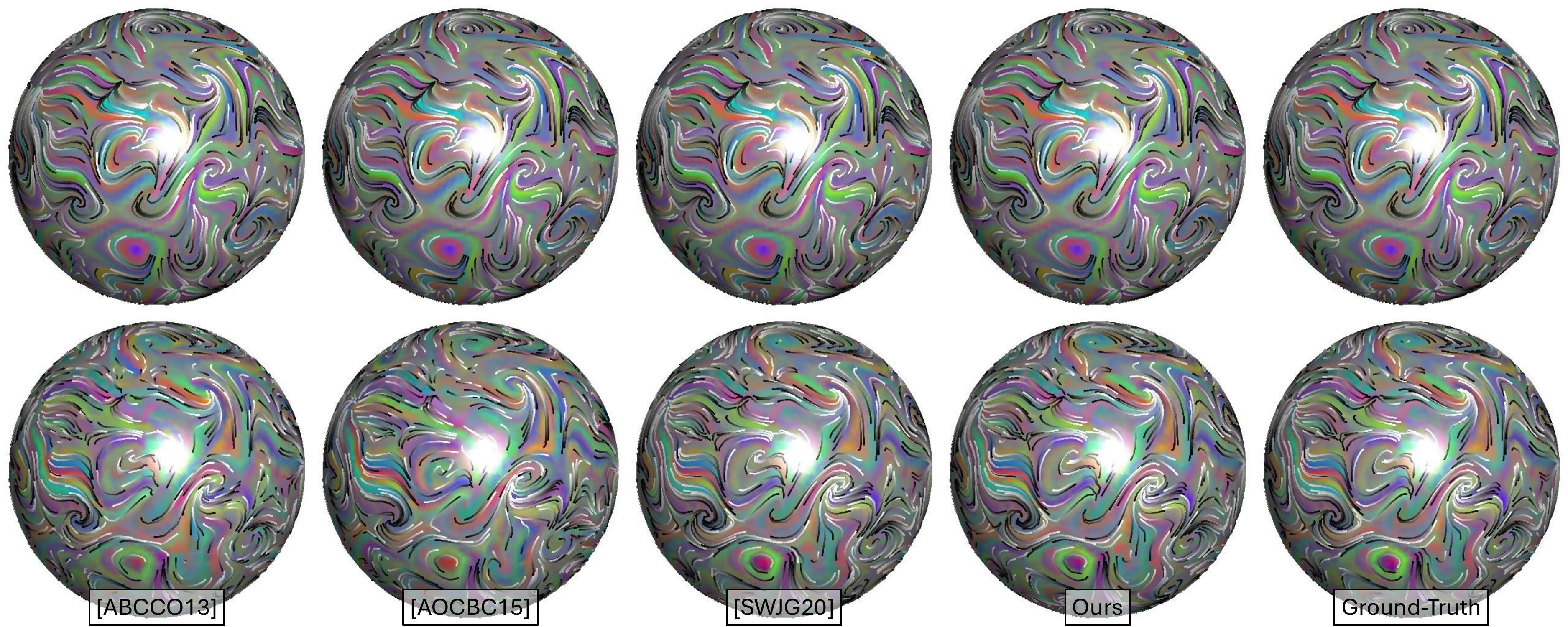}
\end{center}
    \caption{Visualization of the brackets computed for a frequency $b=10$ function using a regular tessellation obtained by subdividing an icosahedron (top) and an irregular tessellation obtained by randomly sampling points on the unit sphere and computing the convex hull (bottom).}
    \label{fig:bracket_compare}
\end{figure*}

\fig{fig:bracket_compare} visualizes the results, showing the bracket estimated using the four different approaches (first four columns) as well as the analytic solution (right), for vector-fields $\tilde{X}$ and $\tilde{Y}$ with band-width $b=10$. On the uniform tessellation of the sphere obtained by subdividing the icosahedron (top) all methods match the ground-truth. However, on the non-uniform tessellation the brackets computed using the operator-based approaches produce tangent vectors whose magnitudes do not match those of the ground-truth. (Note that both the seeding of the initial noise for anisotropic diffusion, and the seeding of flow-lines at triangle centers are tessellation dependent. As such, while the visualizations in the rightmost column represent the same ground-truth vector-field, they appear different.)

\paragraph*{Performance Under Refinement}
\begin{figure}[tb]
\begin{center}
    \includegraphics[width=\columnwidth]{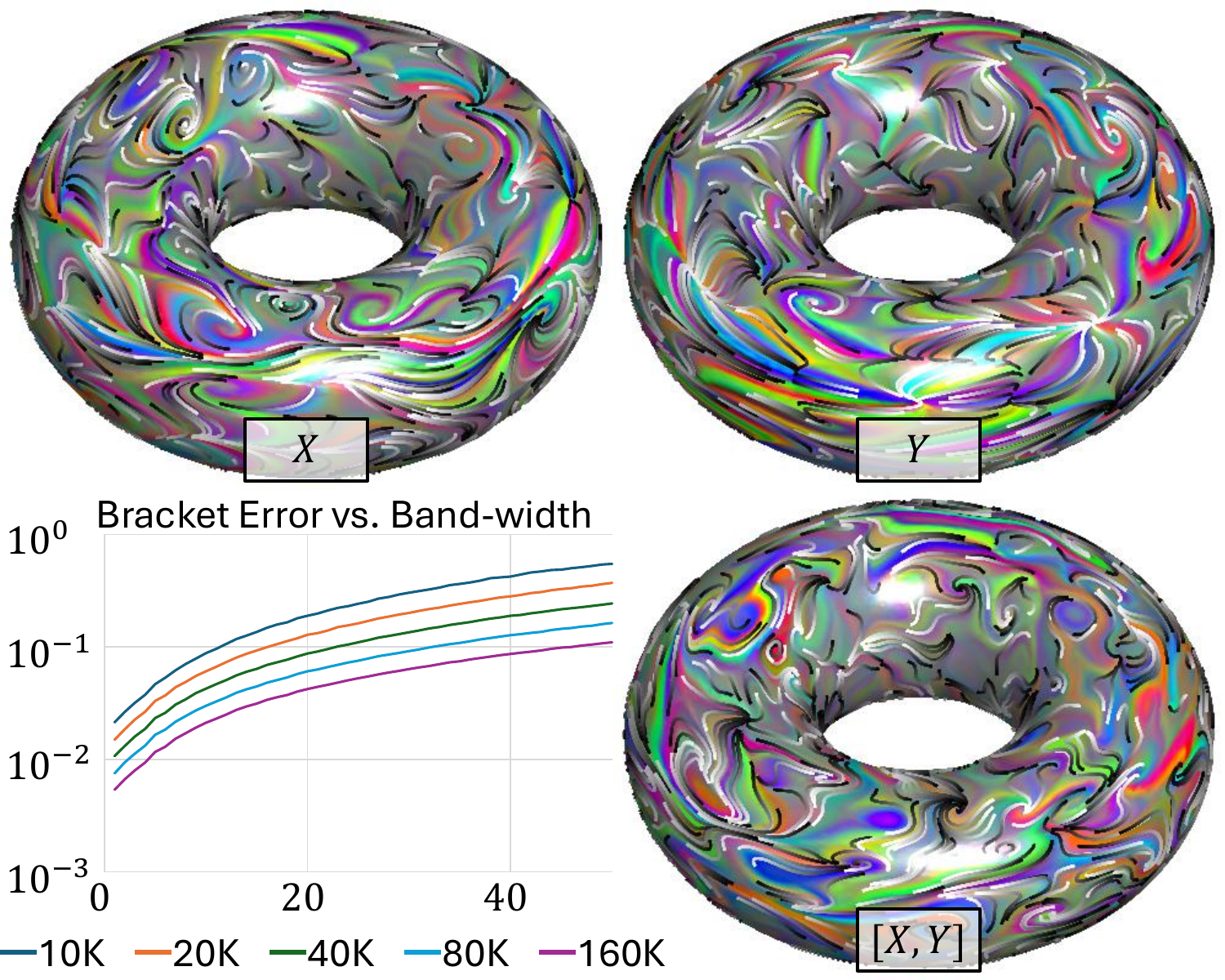}
\end{center}
    \caption{The relative error between the estimated Lie bracket of two vector-fields on a torus and the analytic solution, given as a function of the band-width of the vector-fields. The plots show that error increases with the frequency of the vector-fields, and decreases as the resolution is increased.}
    \label{fig:bracket}
\end{figure}

We also evaluate the performance of our method under refinement. To this end we construct tessellations of the torus by randomly sampling points in the periodic plane, computing the Delaunay triangulation, and mapping to the torus in 3D with inner (resp. outer) radius $2.0$ (resp. $1.0$):
$$\Phi(s,t) = \big( \cos(s),0,\sin(s) \big)\cdot(2+\cos(t)) + \big(0,\sin(t),0\big).$$

As above, we synthesize band-limited vector-fields with uniformly random Fourier coefficients, $X,Y:\R^2\rightarrow\R^2$, this time in the parameterization domain. Leveraging the metric-independence of the Lie bracket, we obtain the ground-truth by analytically computing $[X,Y]:\R^2\rightarrow\R^2$ and using the differential of the parameterization $\Phi$ to map it to 3D. We compare this to the bracket obtained by applying our method to the 3D vector-fields obtained by applying the differential of the embedding, $d\Phi$, to $X$ and $Y$.

\fig{fig:bracket} (bottom left) shows the relative errors between the estimated and ground-truth brackets, computed for varying band-widths $b\in[1,50]$ and averaged over 10 tessellations. As expected, error increases with frequency and decreases consistently as we increase the resolution of the tessellation from $10K$ vertices to $160K$. The figure also shows an example of two vector-fields, $X$ and $Y$, computed at a bandwidth of $b=10$ using $80K$ vertices (top) and the estimated Lie bracket $[X,Y]$ (bottom right).

%% file: Sections/08_discussion.tex
%\subsection{Properties and Limitations}
%\paragraph*{Choice of normals} Our construction defines a discretization using the Phong interpolation of per-vertex normals. Implicit in this approach is the assumption that the interpolated normals are consistent with the triangle normal (e.g. they point in the same direction). For all our evaluations we defined per-vertex normals by summing the area-weighted normals of incident faces.

\subsection{Future work}
Our work is distinct from other directional-field methods in that it decouples the tangent bundle from the geometry of the mesh. We demonstrated that this unlocks a potential for smooth fields with PL vertex-based normals, and we believe it has the potential for more flexible choices of tangent bundle representation.
In the immediate future, we would like to consider extensions of our work using second-order Lagrange elements for discretization, and using a hierarchy akin to that in Subdivision Exterior Calculus~\cite{10.1145/2897824.2925880} to support computation over a smoother surface. More broadly, we would like to explore incorporating structure preservation into our discretization by building cochain sequences and optimizing the choice of per-vertex normals to improve the quality of the discretization.
Additionally, it may be possible to extend our work to vector-field processing with general normal maps~\cite{tasdizen2003geometric}, allowing for detailed field editing on coarse meshes. \revision{Finally, we would like to develop polar representations~\cite{Crane:2010:TCD, Wang2025} for our fields, to be able to explicitly control singularities.}

\subsection{Discussion}
\paragraph*{Rotation} Like the discretizations of Kn\"oppel~\etal~\cite{Knoppel:2013:TOG} and Stein~\etal~\cite{Stein:2020:CGF}, assuming that the triangle mesh is orientable, our discretization is closed under the action of counter-clockwise rotation by $90^\circ$ in the tangent plane. In particular, this suggests that the discretization does not exhibit a preference for divergence-free vs. curl-free vector-fields. This is in contrast to the Whitney 1-form basis that only exhibit curl in the interior of triangles, and whose $90^\circ$ counter-clockwise rotations are not in the space of Whitney 1-forms.

\paragraph*{Relation to Scalar Gradients} As with the discretization of Kn\"oppel~\etal~\cite{Knoppel:2013:TOG}, our vector-fields are continuous across the mesh. An implication of this is that they cannot represent the gradients of functions expressed in terms of the hat-basis, as those gradients are piecewise constant. %However, it is straight-forward to incorporate the gradients of scalar functions within the vector-field-processing construction. (For example, one can integrate the gradients against the continuous basis vector-fields to get a ``weak'' representation that can then be used to find the projection onto the space of continuous vector-fields by solving with the mass matrix.)

\paragraph*{Implicit Geometry}
In our discretization, we implicitly treat the defined $\widetilde{d\Phi}|_p$ operator as the differential of an embedding, using it to define the mapping from the tangent space of the unit right triangle to the tangent space of the sphere. However, there may not be any $\Phi$ whose differential is $\widetilde{d\Phi}$. For example, the rows of $\widetilde{d\Phi}|_p$ may not be curl-free. One immediate implication is that the associated second fundamental form, $\widetilde{d\Phi}|_p^\top \cdot dN|_p$, need not be symmetric. It would be interesting to consider other choices for $\widetilde{d\Phi}|_p$, including ones that do not define an orthogonal transformation between $T_p\T$ and $T_{N(p)}S^2$, thereby inducing a non-constant Riemmanian metric within each triangle.

%In the future, we would like to explore addressing this limitation in one of two ways: (1)~optimizing the per-vertex normals and (2)~optimizing the definition of $\widetilde{d\Phi}|_p$. Like the approach of Boubeker~\etal~\cite{Boubekeur:2008:TOG}, the latter approach could seek the geometry that fits the prescribed normals.

\begin{comment}
\paragraph{Future Work}
\begin{itemize}
\item Extend to frame field processing.
\item Extend to second-order Lagrange elements.
%\item Extend to subdivision exterior calculus.
\end{itemize}
\end{comment}

%% file: Sections/A3_appendix.tex
Given an inner-product space $\{V,B:V\rightarrow V^*\}$, with $B$ symmetric and positive definite, we would like to show that the maps taking an endomorphism $L\in\hbox{End}(V)$ into $\scalar{V}$, $\noscalar{V}$, and $\asym{V}$ are orthogonal projections.

To show that a mapping is a \textit{projection} onto a subspace requires showing (1)~that the image of the map is the subspace and (2)~that the map acts as the identity on the subspace. To show that the projection is \textit{orthogonal} it suffices to show that it is self-adjoint.

We review how inner-products $B_V:V\rightarrow V^*$ and $B_W:W\rightarrow W^*$ on vector spaces $V$ and $W$ define an inner-product on the space of homomorphisms between $V$ and $W$. In particular, this allows us to define an inner-product on the space of endomorphisms $\hbox{End}(V)$, which is needed to show that a projection is orthogonal.

Then, we step through the derivations showing that the map onto the subspace of self-adjoint operators:
$$L\mapsto \left(\frac{L+B^{-1}\circ L^*\circ B}2\right)$$
is an orthogonal projection. (A similar argument shows that the other maps are orthogonal projections as well.)

\subsection*{Inner-Products on Linear Maps}
Given inner-product spaces $\{V,B_V\}$ and $\{W,B_W\}$, there is a canonical inner-product on the space of linear maps between $V$ and $W$, defined in terms of the trace of an endomorphism on $V$. Concretely, for linear maps $L,M\in\hbox{Hom}(V,W)$:
\begin{equation}
\label{eq:hom_inner_product}
\langle L,M\rangle
\equiv\hbox{tr}\left(B_V^{-1}\circ L^*\circ B_W\circ M\right).
\end{equation}
%with $L^*:W^*\rightarrow V^*$ the dual map.
This definition is independent of the choice of bases for $V$ and $W$. We note that the argument to the trace is itself an endomorphism on $V$. Concretely: $M$ maps $V$ to $W$; $B_W$ maps $W$ to $W^*$; $L^*$ maps $W^*$ to $V^*$; and $B_V^{-1}$ maps $V^*$ back to $V$. Thus, the trace is well-defined. 

In particular, when operators are represented w.r.t. orthonormal bases on $V$ and $W$, so that the inner-products $B_V$ and $B_W$ are represented by the identity matrices, the expression reduces to
$$\hbox{tr}(\mathbf{L}^\top\cdot\mathbf{M})\equiv\langle\mathbf{L},\mathbf{M}\rangle_F,$$
the Frobenius inner-product of the matrices $\mathbf{M}$ and $\mathbf{L}$ associated to $M$ and $L$ respectively.

\subsection*{Orthogonal Projection}
We begin by showing that the map:
\begin{align*}
\pi:\hbox{End}(V)&\rightarrow\hbox{End}(V)\\
L&\mapsto\frac{L+B^{-1}\circ L^*\circ B}2
\end{align*}
is a projection onto the subspace of self-adjoint endomorphism and then show that it is orthogonal.

\subsubsection*{Projection}
To show that $\pi$ is a projection we show that its image lies in the subspace of self-adjoint operators and that it acts as the identity on that subspace.
\paragraph*{Image of $\pi$}
%We recall that an endomorphism $L\in\hbox{End}(V)$ is self-adjoint if $B\circ L=L^*\circ B=(B\circ L)^*$.
Noting that that the dual of the composition is the composition of the dual in reversed order (i.e. $(L\circ M)^*=M^*\circ L^*$) and that $B$ is symmetric (i.e. $B=B^*$ and $(B^{-1})^*=B^{-1}$):
\begin{align*}
B\circ\pi(L)
&=B\circ\left(\frac{L+B^{-1}\circ L^*\circ B}2\right)=\left(\frac{B\circ L+L^*\circ B}2\right)\\
&=\left(\frac{L^*\circ B+B\circ L}2\right)^*
%=\left(B\circ\left(\frac{B^{-1}\circ L^*\circ B+ L}2\right)\right)^*\\
%&=\left(L+\frac{B^{-1}\circ L^*\circ B}2\right)^*\circ B=\pi(L)^*\circ B.
=\big(B\circ\pi(L)\big)^*=\pi(L)^*\circ B.
\end{align*}
Thus, the image of $\pi$ is self-adjoint.

\paragraph*{Identity on the subspace} In the case that the endomorphism $L$ is self-adjoint -- i.e. that $B\circ L = L^*\circ B$ -- we have:
\newcommand{\smallcirc}{\raisebox{0.25ex}{$\scriptscriptstyle \circ$}}
$$
\pi(L)=\left(\frac{L+B^{-1}\circ L^*\circ B}2\right)=\left(\frac{L+B^{-1}\circ B \circ L}2\right)=
%\left(\frac{L+L}2\right)=
L.
$$

\subsubsection*{Orthogonality}
To show that $\pi$ is an orthogonal projection, we show that it is self-adjoint. That is, for all $L,M\in\hbox{End}(V)$ we show that:
$$\langle\pi(L),M\rangle=\langle\pi(M), L\rangle,$$
with the inner-product on endomorphisms induced from the inner-product on $V$ (\eqn{eq:hom_inner_product}). Expanding twice the left-hand-side, we get:
\begin{align*}
2\cdot\langle\pi(L),M\rangle
&=\langle L,M\rangle+\langle B^{-1}\circ L^*\circ B,M\rangle\\
&=\langle L,M\rangle+\hbox{tr}\left(B^{-1}\circ\left(B^{-1}\circ L^*\circ B\right)^*\circ B\circ M\right)\\
&=\langle L,M\rangle+\hbox{tr}\left(B^{-1}\circ\left(B\circ L\circ B^{-1}\right)\circ B\circ M\right)\\
&=\langle L,M\rangle+\hbox{tr}\left(L\circ M\right).
\end{align*}
As this is symmetric in $L$ and $M$, the projection $\pi$ is self-adjoint.

%% file: Sections/A2_appendix.tex
\begin{table}
\small
\center{
  \begin{tabular}{r|ccc@{}}
  & \cite{Sharp:2019:TOG}
  & \cite{Stein:2020:CGF}
  & Ours
  \\
\midrule
Without unfolding& $0.335$ & $0.468$ & $\textbf{0.332}$ \\
With unfolding & $0.214$ & $0.222$ & $\textbf{0.208}$
\end{tabular}
%\begin{tabular}{@{}r@{\,\,}l|ccc@{}}
%  &
%  & \cite{Whitney:GIT:1957}
%  & \cite{Stein:2020:CGF}
%  & Ours
%  \\
%\midrule
%Hodge &            & $0.773$ & $0.458$ & $\textbf{0.324}$
%Hodge & (unfolded) & $0.595$ & $0.187$ & $\textbf{0.176}$ \\
%  \end{tabular}
  \caption
	{
    Relative smoothness of the interpolating vector-fields obtained using a smoothness energy defined by the Connection Laplacian, and visualized in \fig{fig:interpolation} (top).
    \label{tab:sparse_interpolation}
  }
}
\end{table}
Although a ground-truth solution for the sparse interpolation problem described in \sec{ss:example_applications} is not available, one can consistently measure the smoothness of the different results. To this end, we sample the different solutions at the faces to obtain a per-triangle representation of the vector-field, and measure the difference between vectors assigned to triangles on opposite sides of an edge.

For our approach, the per-triangle vectors are obtained by directly evaluating the basis functions to obtain extrinsic vectors. For the methods of Stein~\etal and for the Whitney basis, this is done by first evaluating the basis functions at the centers of the triangles to obtain an intrinsic representation of the tangent vectors, and then using the triangles' embeddings to obtain extrinsic vectors tangent to the triangles. For the method of Sharp~\etal, we interpret the per-vertex coefficients as extrinsic vectors perpendicular to the vertices' normals and compute their average.

We measure the relative smoothness by taking the sum of squared differences between the vectors associated to triangles incident on an edge, weighted by the ratio of the (absolute) primal-to-dual edge lengths. The error is normalized by the area-weighted magnitude of the vector-field.

We compare the vector-fields in two ways: (1)~We directly compute the weighted sum of squared differences; and (2)~Before computing the contribution of an edge, we first perform a ``hinge-unfolding'' to bring the two incident triangles into a common plane and compare the aligned vectors. Because our method does not generate tangent vectors perpendicular to the triangle's normal, we first apply the Rodrigues rotation taking the interpolated normal to the triangle's normal before performing the hinge-unfolding. Similarly, for the method of Sharp~\etal, we obtain vectors perpendicular to the triangle's normal by applying the Rodrigues rotation mapping the vertex normals to the triangles' normals, before averaging to get the per-triangle vector. 

\tab{tab:sparse_interpolation} gives the relative smoothness for the different methods visualized in \fig{fig:interpolation} (top), indicating that our method consistently generates a smooth interpolant. In addition, we find that the hinge-unfolding provides a lower smoothness energy for all methods, which is expected as vectors are not forced to point in different directions when the associated triangles meet at a sharp edge. 

As the smoothness mimics the Connection Laplacian -- e.g. it associates a non-zero energy to harmonic vector-fields -- we do not believe it provides a meaningful measure of the quality of the interpolated field. (However, in this case as well, the relative smoothness for the vector-field computed using our is method is lower than that computed using either the Whitney basis or the method of Stein~\etal)

%Finally, we note that the empirical results for Hodge interpolation should be taken with a grain of salt, as the measure of relative smoothness we use mimics the connection Laplacian and penalizes harmonic vector-fields, despite the fact that they have no energy from the perspective of the Hodge Laplacian.

%We note that the discretization of Sharp~\etal is identical to that in the work of Kn\"oppel~\etal~\shortcite{Knoppel:2015:SPS}. We had also compared to the earlier work of Kn\"oppel~\etal~\shortcite{Knoppel:2013:TOG} and found the interpolation results to be very similar to those of their later work.

%% file: Sections/A1_appendix.tex
%As noted by Stein~\etal, since their expression of the covariant derivative that can be evaluated point-wise, one can use the factorization of endomorphisms (\sec{ss:linear_algebra}) to define a Killing energy.

\fig{fig:stein} compares the smallest eigenvectors obtained using our discretization of the Killing energy (first row), with the eigenvectors obtained using Stein~\etal's discretization. As noted in that work, directly using the Killing energy defined by their discretization is not robust due to spurious minimizers. They address this by adding an additional connection energy regularizer to the Killing energy. The different rows show the eigenvectors obtained using Stein~\etal's discretization for progressively larger values of regularization weight $\alpha$. The visualization also gives the associated eigenvalue in the top left corner of each figure.

While a small regularization weight $\alpha=10^{-4}$ effectively generates the as Killing-as-possible vector-field associated with the near rotational symmetry of the pulley, the problem with spurious minima is evident in subsequent eigenvectors. This is mitigated by increasing the regularization weight to $\alpha=10^{-3}$, though even in this case one sees artifacts in the right-most column. While further increasing the regularization weight to $\alpha=10^{-2}$ produces results similar to ours, this can have the detrimental effect of biasing the spectral decomposition away from Killing vector-fields in preferences of vector-fields that are smoother.

This highlights a limitation observed by Stein~\etal~-- while their discretization provides properties desirable for the connection Laplacian (e.g. linear reproduction), the underlying Crouzeix-Raviart basis is discontinuous and can result in artifacts when used to define other energies that depend on vector-field derivatives.